\documentclass{article}
\usepackage[utf8]{inputenc}
\usepackage{graphicx}
\usepackage[varg]{txfonts}
\usepackage[utf8]{inputenc}
\usepackage{subcaption}
\usepackage[breaklinks=true]{hyperref}
\usepackage{graphicx}
\usepackage{amssymb}
\usepackage{svg}
\usepackage{authblk}

\usepackage{titlesec}
\titleformat{\paragraph}
{\normalfont\normalsize\bfseries}{\theparagraph}{1em}{}
\titlespacing*{\paragraph}
{0pt}{3.25ex plus 1ex minus .2ex}{1.5ex plus .2ex}

\title{US ATLAS and US CMS HPC and Cloud Blueprint}

\author[1]{Fernando Barreiro Megino}
\author[2]{Lincoln Bryant}
\author[3]{Dirk Hufnagel}
\author[4]{Kenyi Hurtado Anampa}
\affil[1]{University of Texas at Arlington}
\affil[2]{University of Chicago}
\affil[3]{Fermilab}
\affil[4]{University of Notre Dame}

\date{November 2022}

\begin{document}

\maketitle

\begin{abstract}

The Large Hadron Collider (LHC) at CERN houses two general purpose detectors - ATLAS and CMS - which conduct physics programs over multi-year runs to generate increasingly precise and extensive datasets. The efforts of the CMS and ATLAS collaborations lead to the discovery of the Higgs boson, a fundamental particle that gives mass to other particles, representing a monumental achievement in the field of particle physics that was recognized with the awarding of the Nobel Prize in Physics in 2013 to François Englert and Peter Higgs. These collaborations continue to analyze data from the LHC and are preparing for the high luminosity data taking phase at the end of the decade. The computing models of these detectors rely on a distributed processing grid hosted by more than 150 associated universities and laboratories worldwide. However, such new data will require a significant expansion of the existing computing infrastructure. To address this, both collaborations have been working for years on integrating High Performance Computers (HPC) and commercial cloud resources into their infrastructure and continue to assess the potential role of such resources in order to cope with the demands of the new high luminosity era. US ATLAS and US CMS computing management have charged the authors to provide a blueprint document looking at current and possibly future use of HPC and Cloud resources, outlining integration models, possibilities, challenges and costs. The document will address key questions such as the optimal use of resources for the experiments and funding agencies, the main obstacles that need to be overcome for resource adoption, and areas that require more attention.

\end{abstract}

\tableofcontents

\section{Executive Summary}
ATLAS and CMS are the two general purpose detectors at the Large Hadron Collider at CERN. The processing unit is the collision of two particle bunches (usually referred to as event), which can be processed by a CPU core and some GB of RAM within seconds. The workloads are trivially parallelizable and the computing complexity lies in the amount of collisions that need to be processed and simulated. The experiments' computing models rely on a worldwide distributed processing grid of computing centers hosted by associated universities and laboratories. The grid is optimised for loosely-coupled tasks, runs the whole year through and delivers a roughly constant CPU power.

Other widely available computing infrastructures nowadays are HPC centers for large, highly complex workloads, and commercial clouds with a pay-as-you-go pricing model. ATLAS and CMS have gathered significant experience with these resources in the last decade and integrated them with their frameworks alongside dedicated grid resources. US ATLAS and US CMS computing management has charged the authors (see \ref{charge}) to write a blueprint document describing integration models, possibilities, challenges and cost. What is the best usage of these resources for the experiments and for the funding agencies? What are the main difficulties that need to be tackled in order to simplify the adoption of the resources? Which areas need more effort?

Today we are well-positioned to utilize many HPCs for a few reasons: \begin{itemize}
    \item Current generation HPCs have largely settled on x86 architecture, the same as grid sites, making it more straight-forward to run software there as all of the existing physics validation and software packaging has been done for this architecture
    \item Operating system advancements have made it much easier to bring experiment software, conditions data, etc to the compute nodes via containerization, squid caching infrastructure and CVMFS
    \item Data transfer mechanisms and performance to/from the wide-area network are good enough at the scale of computing and facilities used for today's workloads
\end{itemize}

However, there are still significant challenges, and a number of opportunities, to make the best possible use of HPC resources. Among these are: \begin{itemize}
    \item Given the significant recent industry-wide interest in Artificial Intelligence and Machine Learning (AI/ML) via GPU acceleration, as well as the general performance-per-Watt benefits of accelerators, it is likely that HPC facilities of all sizes and scopes will increasingly build new resources around accelerator technologies. It is therefore important for ATLAS and CMS to commit to more R\&D around GPU technology, whether it is porting existing codes to accelerators or taking advantage of AL/ML techniques.
    \item While HPCs provide significant computing resources to ATLAS and CMS, the storage requirements of the collaborations need to be considered as well. Today HPC storage is used primarily to stage transient data to and from the respective data federations used by ATLAS and CMS. Grid sites, in turn, must provide more storage overall to make up for the lack of long-term storage at HPCs. The collaborations should work together and with HPC providers to develop a plan for providing large allocations for storage that can federate with existing ATLAS/CMS data management systems. 
    \item Many HPC facilities are beginning to investigate, or are already supporting, Kubernetes-based edge service platforms that sit adjacent to supercomputer resources. These platforms give users the ability to run long-lived services that can interact with the HPC in some way. While this technology is nascent, ATLAS and CMS should investigate the feasibility of running their respective HPC resource integration technologies on these edge platforms.
\end{itemize}

Advancements in these areas will help to make better use of HPC in general, but especially with the Leadership Class Facilities (LCF). The LCF present the best opportunity for growing our HPC usage, both because we currently don't use them much and also because they are by far the largest HPC resource available. 

As our colleagues in Europe have shown with the Vega supercomputer, it is possible for HPCs to handle a significant fraction for LHC computing needs if the experiments are involved in the early phases of machine design. 
To that end, ATLAS and CMS should engage the HPC facilities and together with them document the experiments' requirements and desiderata for future supercomputer designs. This document should also include the need for multi-year compute and storage allocations such that ATLAS and CMS can incorporate HPC resources into their planning horizons. 

% Multiyear allocations, planning, whitepaper.

On the other hand, major commercial clouds offer the building blocks to configure and operate a grid equivalent site with the required compute and storage services. The effort goes into the setup and validation of the resources in order to get a reliable infrastructure and operate it in the most cost-effective manner. Once in operation the infrastructure is mostly autohealing and autoscaling, requiring very little maintenance. 
From a technical point of view, the cloud offers a very powerful substrate and we we will give examples for different use cases that we have exercised in the cloud:
\begin{itemize}
    \item Bulk processing: running a grid-equivalent site with storage and compute. We have ran it both in a flat, medium-scale scenario and a bursty, large-scale scenario. 
    \item Infrastructure for user analysis: we have set up a facility for physics analysis allowing end-users to spin up private, on-demand clusters for parallel computing with up to 4000 cores, or run GPU enabled notebooks for machine learning applications.
    \item Usage of special hardware: we have used the cloud to make available and validate special hardware, which is not widely available at on-premise resources. Examples are single FPGA, TPU or very large nodes, as well as ARM and GPU clusters.
\end{itemize}

There are many other services and platforms that can still be explored and integrated. 

Currently, the major questions around the cloud are the cost of the resources and long-term governance.
\begin{itemize}
    \item We need to find a balance and the most cost-effective cloud usage. The main concerns are around the egress costs (ie. charges for outgoing network traffic from the Cloud) .
    \item What is the role of the cloud within the WLCG? What is the sweet spot that provides the biggest advantages for the money spent?
    \item What data can be trusted to the cloud? Can we rely on the cloud to store unique copies of data?
    \item How can we shield us contractually in the long term regarding cost and to protect our data?
\end{itemize}
We will provide cost estimations or extracts from the billing reports for the different use cases evaluated above. We will also explain the attractive subscription model that ATLAS engaged on with Google cloud, but for which we do not have experience in the renegotiation of terms.

There is still work remaining for a full integration of HPC and Cloud resources in the WLCG. If HPC or Cloud not only wants to be considered an opportunistic or beyond-pledge option, there are discussions whether these resources comply with some of the memorandum of understanding requirements, in particular providing a constant amount of compute power and being able to execute any type of workload. The pillars for pledging need to be addressed first. These are benchmarking (measuring the power of the CPUs) and accounting (accounting and reporting the CPU usage). 
\begin{itemize}
    \item Benchmarking will be complicated but necessary to compare the relative computing power of individual cloud/HPC resources. 
    \item An alternative method to report the accounting numbers to WLCG needs to be implemented when bypassing the Compute Element and connecting the batch system directly to the experiment’s Workload Management System. Solutions should be extended to a generic service, where resource-specific or Workload Management System based plugins need to be added.
\end{itemize}    

Both experiments' frameworks were originally built for only CPU resources. Over the last few years CMS has included GPU support and is using (Nvidia) GPU resources in the High Level Trigger since the start of Run3 in 2022, with the ability to use GPU resources for (some) offline production workflows planned for 2023. ATLAS is making use of GPU resources so far mainly in machine learning applications for user data analysis. While there are no ATLAS production workflows running on heterogeneous computing systems yet, accelerator resource management and the optimization of hybrid CPU/GPU workflows is the highest R\&D priority for the core software team targeting the HL-LHC by porting parts of the tracking reconstruction or detector simulation codes to efficiently use GPUs.
Independent of the technical ability to run workflows that use GPU, larger scale adoption of such resources is hindered by a lack of experience with GPU benchmarking and accounting (and related also the inability to pledge such resources).

We will also discuss some open points around infrastructure integration, for example the inclusion of commercial cloud Certificate Authorities into the WLCG trust model and connecting HPCs and commercial clouds with the WLCG science data network.

To conclude, we made six recommendations and proposed twelve common R\&D directions summarized in sections \ref{comments-recommendations} and \ref{common-rnds}.

\section{Introduction}

Particle physics is the study of the most elementary components of matter and their interactions. As such, it addresses existential questions about the universe. The key experimental tools of the field are particle accelerators, which bring everyday particles such as protons into high-energy collisions. These collisions occur at energy scales representative of the very early universe. By replicating the conditions of the universe in its infancy, we can study the interactions of particles under those conditions and thereby understand how today's universe came to be.

%Run3 will be at 12.6TeV, need to change language?

The Large Hadron Collider (LHC) is currently the world's highest-energy accelerator. It collides protons (and, for certain special runs, heavy ions) at a center-of-mass energy ($\sqrt{s}$) of about 13~TeV. The collision energy can be used to create heavy particles that existed in the early universe but are rarely observed in our world today. This very high collision energy allows the possibility of producing particles that have never been seen before, giving the LHC great opportunities for discovery.

ATLAS (A Toroidal LHC Apparatus)~\cite{bib:2008zzm} and CMS (Compact Muon Solenoid)~\cite{bib:2008zzk} are the two general-purpose particle physics detectors at the LHC. Both collaborations discovered the Higgs boson in 2012~\cite{ATLAS:2012yve}~\cite{CMS:2012qbp}, have provided constraints on many models of new physics, and have made many precise measurements of the properties of known particles.

Both collaborations are actively analyzing data recorded during ``Run~2" of the LHC from 2016 through 2018, with an emphasis on measurements of the properties of the Higgs boson and searches for dark matter, supersymmetric particles and exotic Higgs bosons. The collaborations are also working on the new ``Run~3" (data taking started in 2022). The prompt delivery of scientific results depends on many elements, with the actual physics analysis only the final step. It starts with data taking, i.e. the recording of the collision events. We also need to run alignment and calibration workflows that give us a better understanding of the detector and feed directly into the data reconstruction. The full suite of ATLAS and CMS physics measurements also requires the simulation of billions of proton-proton collision events, including both processes that would arise from new, speculative physics models and those that arise from standard model processes that would be the background to the new physics.

\subsection{Experiment Workflows}
% what are HEP Workflows
% Not only read by ATLAS/CMS computing people. We need to explain the workflows.
% Write for a general computing audience who are not necessarily familiar with HEP computing.
% e.g.: on ATLAS, generation/simulation, on CMS Reco

%need to check numbers! either update or leave out!

The LHC collides bunches of protons (with $10^{11}$ protons in each bunch) at a rate of 40~MHz. Only a tiny fraction of these collisions are of scientific interest; for instance, Higgs bosons are produced in only about one out of every $10^{10}$ collisions. Still, in any given beam interaction, about 40 protons collide, usually with low momentum transfer. These \textit{pileup} interactions are superimposed on top of any interesting hard scatter that might occur in a single beam crossing. A trigger system performs fast pattern recognition to select collision events of interest for later, offline analysis, reducing the data rate to approximately 1~kHz. The resulting dataset is very general in nature, with a mix of contributing physics processes. Smaller teams of physicists within the CMS and ATLAS collaborations, each with their own scientific interests and targeted measurements, apply their own individual set of selection criteria to further reduce these datasets into samples that are enriched in events of interest to them. But even then, the signal-to-noise ratio can be quite small, in the parts per thousand for measurements that seek to observe new particles that are only produced rarely.

Since data alone cannot be used to infer the physics composition of the event sample, physics measurements rely on detailed simulations of the processes that contribute to the selected samples. Most events come from well-known physics processes, but from very particular corners of their phase space that might not be well understood. These simulations have multiple steps. First, the actual physics process must be simulated. In this \textit{Generation} step, the incoming protons are described and the momenta of new outgoing particles are determined from basic particle physics theories. The list of particle momenta are the input to the \textit{Simulation} step. The outgoing particles are propagated through a model of the material of the detector, and undergo energy loss and scattering according to established models and a description of the detector geometry. How the active detector elements, including the readout electronics, respond to the energy losses is part of the \textit{Digitization} step, which also includes running the trigger software on the simulated event and the superposition of additional $\sim 40$ soft proton collisions to simulate the pileup. The output of that step is in a format that is completely identical to the recorded data from the actual detector. The final step is then \textit{Reconstruction}, in which pattern recognition algorithms are applied to the recorded and simulated detector outputs to determine the momenta and energies of the particles observed in collision events. This reconstruction step is the same one that is also applied to the recorded data from the actual detector.

The first three steps (all but \textit{Reconstruction}) are done through Monte Carlo calculations. \textit{Generation} uses codes such as PYTHIA~\cite{bib:pythia} and Madgraph~\cite{bib:madgraph}, \textit{Simulation} relies heavily on the GEANT particle propagation software~\cite{bib:geant}, and \textit{Digitization} is performed with experiment specific code developed by ATLAS and CMS. All of these codes are integrated into the experiments' software stacks, referred to as Athena ~\cite{bib:athena} for ATLAS and CMSSW~\cite{bib:cmssw} for CMS. In addition, the entire simulation process is pleasingly parallel. Each collision event is statistically independent of the others, and thus each one can be simulated and reconstructed independently, with the completed events output concatenated at the end.

\begin{figure}
\centering
\includegraphics[scale=0.2]{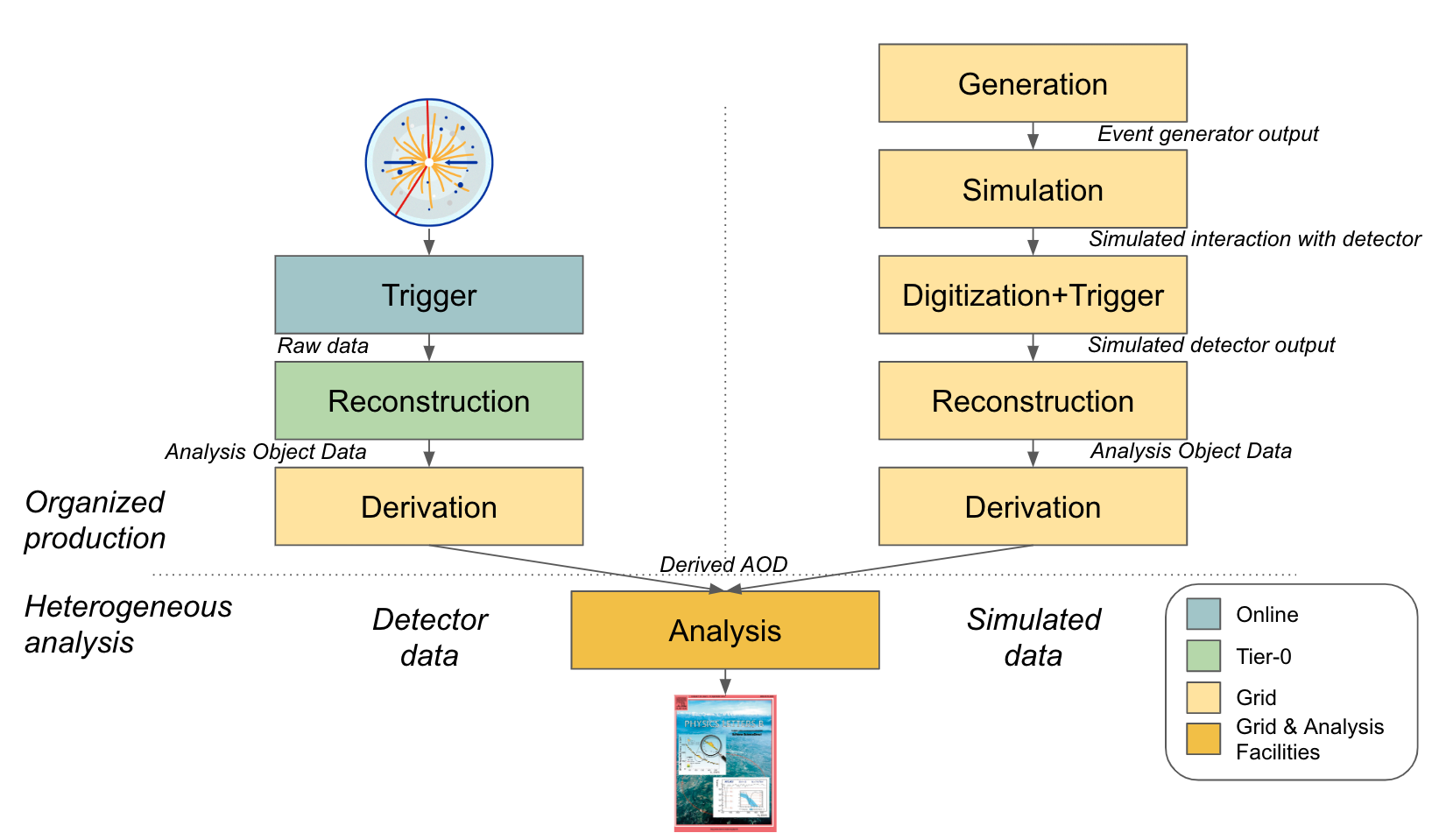}
\caption{Workflow overview in the ATLAS and CMS experiments.}
\label{fig:workflow_atlas}
\end{figure}

Figure~\ref{fig:workflow_atlas} shows an overview of the processing chain and the workflows contained therein for the ATLAS experiment. For CMS this would look very similar, although the terms might be different for some of it (\textit{MiniAOD} and \textit{NanoAOD} production instead of \textit{Derivation} for instance).

%do I need a figure for CMS as well?

The experiment software and other middle-ware are installed and distributed through the CernVM File System (CVMFS)~\cite{bib:cvmfs}. CVMFS is a read-only file system designed to deliver scientific software on to the worker nodes through a hierarchical distribution network similar to a content delivery network. The installation of CVMFS on any compute resource is usually the first integration step.

% Data processing chain: https://docs.google.com/presentation/d/1vWopXz8pmAZD_5hXcQFBbcoXmbAG6NDfLohDYfUk-xg/edit#slide=id.g130bda0d403_0_0

\subsection{The Worldwide LHC Computing Grid (WLCG)}
%some short introduction on our computing models and infrastructure
%including the move from grid-only to include both HPC and Cloud (motivating the next subsections)

% https://inspirehep.net/files/7f22f8d21abc7a812cd2b953b9456f18 - old but good source of information

The LHC experiments' computing models were designed and developed decades ago ahead of the LHC data taking era. At the time, there were no examples in industry or other sciences handling the expected data volumes and computing power. There were no off-the-shelf software solutions that the LHC experiments could use for storage or compute. The WLCG ~\cite{bib:wlcg} was conceived with the purpose to divide the computational challenge across the participating institutes in all member states, and to develop all the necessary software to manage and federate this distributed infrastructure. This distributed model also fitted very well with the collaborative funding model, where all participating countries contribute with a part of the computing resources. National funding agencies prefer to invest locally and promote education, science, technology and industry in their country.

The WLCG has traditionally followed a tiered structure: starting from CERN (Tier 0), going out to national-scale laboratories (Tier 1), then to regional computing centers at Universities or federations of Universities (Tier 2), and finally to small-scale compute clusters in individual PI labs or departments (Tier 3). Historically, the Tier system was a strict hierarchical model with a clear distinction between services and roles hosted at each Tier. 

Over the years, the general IT landscape has evolved very quickly with processing, storage and network capabilities increasing at several orders of magnitude. These advances have been adopted by the WLCG and the strictly hierarchical structure has become a worldwide, highly-connected mesh. The distinction between Tiers has blurred thanks to the excellent facility uptime, performance and network connectivity at all levels. Such improved infrastructure has allowed for example Tier 2 centers to run the full spread of workloads for CMS and ATLAS computing.

\subsection{Where do HPC and Cloud fit in?}

Today, the computational requirements at the LHC are still significant, but the computing world has caught up and in some cases exceeded our needs. LHC computing no longer needs to operate in a bubble and needs to adopt new paradigms with all their advantages and challenges.

In the HPC area, over the last few years there has been a push for Exascale. This year the new OLCF Frontier machine debuted at No.1 on the Top500 list~\cite{bib:top500}, the first machine exceeding 1 ExaFLOPS ($10^{18}$ floating point operations per second) to do so. Historically, HPC weren't used much in HEP since our computing is High Throughput, not High Performance. We process collision events, which represent an individually small amount of compute to process, and since each event is independent from each other event, our computing can be easily parallelized. Therefore we do not need or make use of the fast interconnects available at HPC. Nevertheless, the large capacity increase at HPC has made these machines attractive to us as well.

At the same time, Big Tech companies evolved their business models and started to offer their tremendous compute capacity, technologies and services to customers. Given the ability to pay for it, anyone can rent compute capacity from Amazon Web Services, Google Cloud, Microsoft Azure and others. Many services that used to be developed or operated in-house, can now be directly outsourced, allowing to narrow the focus to the key activities.

Beyond capacity, Cloud and HPC can offer a number of interesting and unique capabilities for HEP computing, including significant Machine Learning and Artificial Intelligence (ML/AI) resources (both in terms of hardware and services) and straight-forward access to new and/or exotic technologies that collaborations would have to otherwise procure themselves. Cloud and HPC projects play a very important role in technology and knowledge transfer, providing new ideas for the experiments software stacks.

Both ATLAS and CMS have explored HPC and Cloud for running their workloads for multiple years now. US ATLAS and US CMS Computing management has therefore charged us (see \ref{charge}) to write a report that gives a comprehensive overview at the current state of integration and usage in the experiments. We focus on the situation and activity in the US, although we have also studied the international status and many aspects will also be relevant globally. Finally, we would like to point out that while this document targets both resources together, it is important to understand that HPC and Cloud are two different resources and have very different characteristics. We will try to provide an outlook on the future and on possible scenarios that benefit the experiments and funding agencies.

\section{High Performance Computing}
\subsection{Landscape of HPC}
\subsubsection{HPC Architecture and System Design}

Historically, HPC resources and WLCG computing applications have not meshed particularly well due to considerable differences in architecture and approaches for delivering data and software to computing resources. This has, in turn, influenced the type of workloads that are sent to HPC centers.

\begin{figure}
\centering
\includegraphics[scale=0.13]{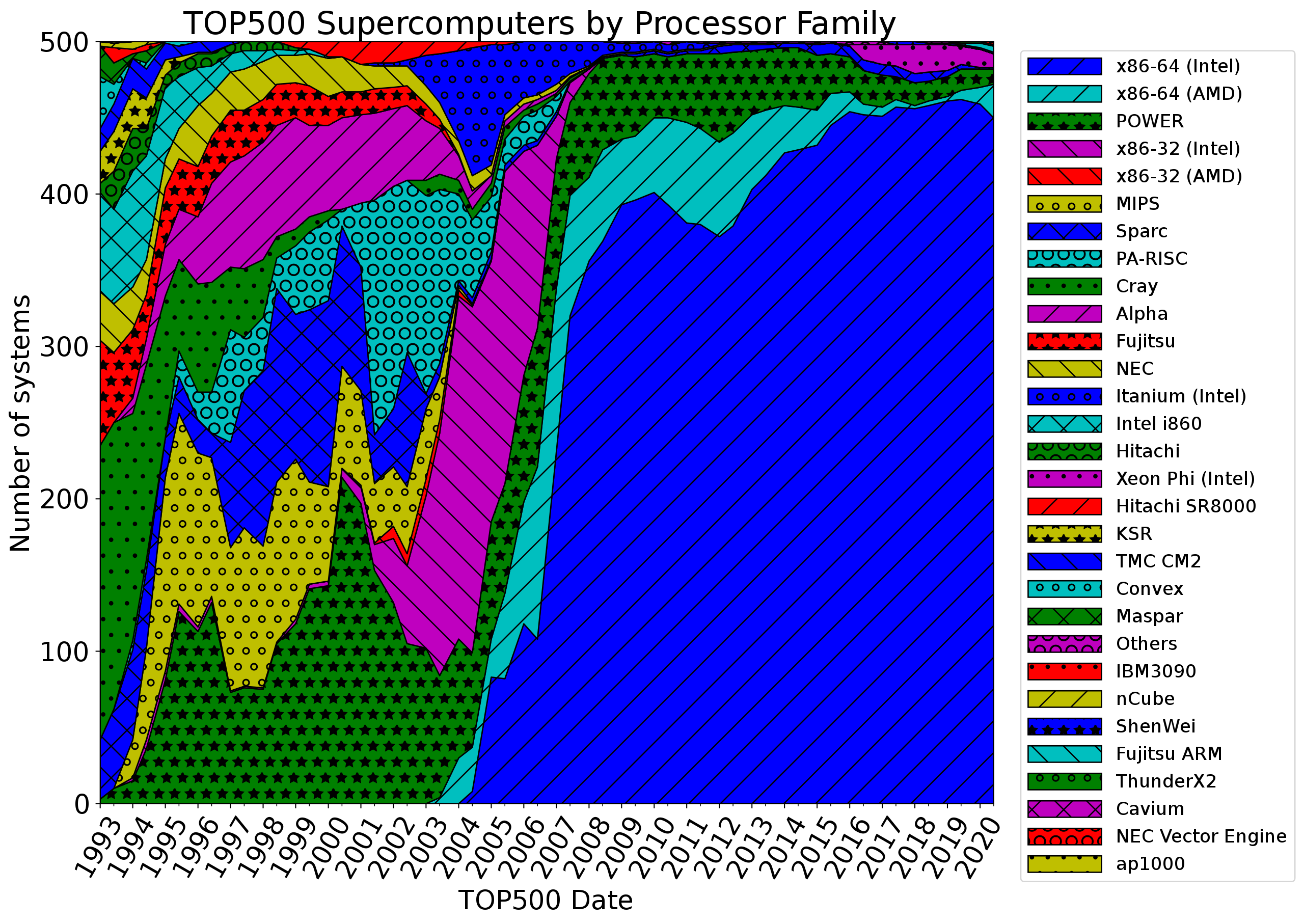}
\caption{Top500 Supercomputers by Processor Family}
\label{fig:top500_processor_family}
\end{figure}

Figure ~\ref{fig:top500_processor_family}~\cite{bib:top500_processor_family} shows that while in the past there was significant diversity in the CPU architectures used at HPC, in recent years \textit{x86-64} (which we will refer to simply as \textit{x86}) based HPC have been dominant.

However, there is no guarantee that this will remain to be the case in the future. IBM is still actively developing the POWER architecture and it might make a comeback. There are plans for new HPC with ARM in Europe and ARM might also be considered for the next generation of US HPC. Particularly interesting are new ARM designs from Nvidia targeted specifically at the AI, HPC, Cloud and Hyperscale market~\cite{bib:nvidia_grace}.

The latter example of the Grace Hopper Superchip, an integrated CPU+GPU design, also points at another development at HPC in recent years. Looking at the current Top500 list, among the first 10 machines by size, only 2 have only CPU. The others are designs with CPU plus GPU/Accelerator, with the vast majority of FLOPS (floating point operations per second) coming from the latter. Talking with HPC facilities about their design considerations for the next HPC systems, this is a likely trend to continue, because GPUs/Accelerators are much more power-efficient in providing a large amount of FLOPS.

One architectural decision made by almost all significant HPC centers is to provide a shared file system for data access. Over the years, we have encountered a number of issues with shared file system performance degradation as a result of overheads from processing many small files. LHC workloads often include many scripts, configuration files, shared libraries, detector environment data and so on, which can cause significant stress to HPC file systems when many jobs start simultaneously. Due to the event processing nature of our computing, our job sizes are also very small compared to HPC norms. We can utilize many nodes at once, but usually only by filling them with many small independent event processing jobs. This exacerbates the stress we put on the shared file system, which affects not only our workloads, but also the workloads of other users on these machines.

Some HPC (not all of them) also provide node local storage, usually flash based. Even though small compared to shared file system quotas, these can be very useful as job scratch space for HEP style workflows, to avoid the aforementioned problems with HEP workflows and shared file systems.

\subsubsection{US HPC Classification}

% Do we want to mention in-house HPCs at labs and universities?
% e.g., Livermore HPC approaches size of XSEDE HPCs - is it worth it to integrate? maybe talk about that later
% maybe just a small paragraph at the end to mention them and say we won't cover them further?

There are broadly two types of HPC facilities in the US, which we will define as Leadership Class Facilities (LCFs) and User Facilities. DOE-funded LCFs push the technological envelope of large scale high-performance computing and are designed for the most demanding tightly-coupled parallel workloads. While these machines have truly exceptional performance, they are not well suited to traditional HEP workloads due to their heavy reliance on GPU accelerator technologies. In addition to the unique architecture of these machines, LCF supercomputers frequently have very restricted wide-area network (WAN) access for performance reasons and to address significant security concerns.

Outside of the Leadership Class Facilities, there are also a number of user-focused facilities that enable considerable large-scale, high-performance computing with more familiar computing architectures and relaxed networking requirements. Facilities such as the National Energy Research Scientific Computing Center (NERSC), the Texas Advanced Computing Center (TACC), and the San Diego Supercomputing Center (SDSC) among others have significant compute resources that are well-suited to HEP workloads in a fashion that is adaptable to interfacing with the traditional grid computing model. 

User Facility HPC currently still provide mostly CPU resources, but this also has begun to change. Perlmutter, the latest HPC deployed at NERSC in 2022, still provides a fraction of CPU-only nodes, but most FLOPS come in the form of nodes with GPU accelerators.

\subsection{HPC Workflows}

Advancements made in the operating system space, in particular Linux containerization mechanisms, are enabling ATLAS and CMS to send a greater variety of workloads to HPCs than in the past. LHC experiments can use Linux containerization wholly via runtime engines such as Apptainer (previously Singularity)\cite{bib:apptainer} or Shifter\cite{bib:shifter}, or in-part via \textit{à la carte} containerization primitives provided by the Linux kernel such as user namespaces. User namespaces also allow the CVMFSExec~\cite{bib:cvmfsexec} tool to provide portable access to CVMFS on sites where it has not been installed system-wide. Along with the current dominance of \textit{x86} architecture in the HPC space, the shift toward containerization has gone a long way toward homogenizing HPC environments and easing the task of porting workflows to these machines.

Although supercomputer CPU architectures are largely compatible with our computing model, an ongoing and persistent challenge with LHC workflows is that HPC facilities are increasingly focusing their machine designs around accelerator technologies to reach the level of FLOPS expected by their user-base. Due to significantly different hardware architectures, taking advantage of GPUs or other accelerators is not as simple as retargeting our software builds for the accelerated platform. As such, HEP software in general will need to be significantly and deeply refactored to take advantage of hardware acceleration where possible. These efforts will likely increase the modularity of HEP software in general, easing possible later transitions to future architectures. While some facilities continue to provide CPU only partitions, we expect this gap will only widen in coming years. Another possibility for taking advantage of accelerated architectures is by the use of novel ML/AI workloads, which will be discussed in later sections.

\subsubsection{ATLAS}

The ATLAS approach to HPCs has been to focus on simulation tasks as they are generally the most likely to run successfully, with the smallest input and output file sizes, and typically the least stressful on HPC file systems. Today, ATLAS submits jobs to HPCs in one of three ways:
\begin{itemize}
    \item If CVMFS is present on the machine, a normal ATLAS pilot job is submitted to the batch system, which can pull down any workflow.
    \item Barring CVMFS, but some network access present, an HPC can run a snapshot of the ATLAS software releases copied locally to the machine
    \item With no CVMFS and no network, ATLAS can send so-called ``fat container" jobs to the HPCs, which contain all software and conditions data needed to run a single release in an isolated environnment
\end{itemize}

Given that many HPCs do not have CVMFS and ATLAS has not yet utilized CVMFSExec, many HPCs require that software releases are copied locally in some manner. In turn, HPC queues without CVMFS require greater maintenance and operational effort as these HPCs require special tasks be created and assigned to them manually rather than through the usual automated process.

As we work toward bringing grid-like capabilities to HPCs and make our workflow management systems friendlier to HPC job execution, we will be able to send a broader variety of workloads to these machines. Success stories such as the EuroHPC Vega supercomputer have shown that it is possible not only to run a large variety of workloads, but also get significant computing output when LHC experiments have early-stage input in the design phases of HPCs.

One of the perennial challenges with executing workloads in an HPC environment is ensuring that job efficiency is reasonably high and that cycles (and therefore allocation) are not wasted. On a resource like NERSC Perlmutter the minimum schedulable resource is 1 compute node, and so a naive approach would be to simply send an Athena simulation job split up into (for example) 256 processes, one for each logical core, per node. While the task will succeed, it is very inefficient (less than 30\% CPU efficiency) due to overheads caused by the initial Athena setup and the merge step for outputs. To overcome some of these inefficiencies, tasks can be arranged such that multiple smaller jobs are co-scheduled on the same compute node, and scheduler parameters at many sites additionally require multi-node batch requests to efficiently fill the machine. While our work is pleasingly parallel, we can again get caught in an inefficient state when a particular simulation takes significantly longer to finish than the rest, as the entire HPC batch job is waiting on this final task to complete (and therefore incurring allocation charges for idle CPUs) before the nodes can be freed for other tasks. 

Another approach for efficiently utilizing HPCs is to take advantage of the ATLAS Event Service to send highly granular tasks to HPCs on an event-by-event level. Raythena\cite{bib:raythena}, for example, is a vertically integrated scheduler for the ATLAS Event Service that uses the Ray task framework to distribute small batches to running Athena processes onto clusters HPC worker nodes and handles merging on the fly. This allows HPCs to be efficiently filled with tasks as well as supporting preemption with minimal loss of work.

% Should we cite some papers here? 
% I Found this one: https://dl.acm.org/doi/pdf/10.1145/3483447
% Doug's paper: https://cds.cern.ch/record/2728398/files/10.1051_epjconf_201921407005.pdf
% https://www.epj-conferences.org/articles/epjconf/pdf/2019/19/epjconf_chep2018_03047.pdf

\subsubsection{CMS} \label{CMS-HPC-Workflows}

The CMS approach in targeting workflows at HPC has always been that while we began with whatever was easiest, we eventually wanted to be able to run the majority of our workflows on HPC and treat them mostly as a standard type of resource that didn't need a lot of special handling. The reason behind this was that we thought this was the only long term maintainable approach, especially considering a potential increase in HPC use within CMS towards the HL-LHC time frame.

One parallel development within CMS Computing over the last years that has helped with this goal is the deployment of StepChain workflows, ie. a workflow where we run the various steps of the event processing chain within the same job and only stageout and keep the final results of the computation. This helps avoid a lot of round-trips of data between compute and managed storage, which has advantages also for our own grid infrastructure, but especially helps on HPC resources.

CMS currently has fully integrated most of the User Facility type HPC in the US (NERSC Cori/Perlmutter, PSC Bridges-2, SDSC Expanse, TACC Stampede2, Purdue Anvil, TACC Frontera) with the CMS computing infrastructure and automatically assigns the majority of its MC workflows to these resources. The work is automatically routed through the HEPCloud portal at FNAL~\cite{bib:hepcloud}, which also handles the resource provisioning at the HPC.

We have either CVMFS available on all the HPC we use or provide it ourselves via the use of CVMFSExec wrappers. Most StepChain MC workflows contain the full chain of MC event processing, namely \textit{Generation}, \textit{Simulation}, \textit{Digitization} and \textit{Reconstruction}. As such, they need access to pile-up data during the \textit{Digitization} step. We provide this through the CMS AAA data federation~\cite{bib:AAA} (in practice this usually means the pileup data is read remotely from FNAL).

While this approach has been very successful for the User Facility type HPC, it is not easily applicable to LCF. The blocking issues there are integration with the HEPCloud portal and the inability to read job input data remotely through the CMS AAA data federation.

\subsection{HPC Integration} \label{hpc-integration}
When ATLAS or CMS submit jobs to typical WLCG grid sites, workflow systems such as PanDA and GlideinWMS~\cite{bib:glideinwms} interface to sites by way of a Compute Element (CE). The Compute Element provides a job gateway abstraction that handles authentication/authorization and routing workloads to local cluster resources at a given site.

HPCs usually do not directly support remote job submission through a CE as it requires a long-lived daemon to be run by the local HPC site administrators, typically on a virtual machine or dedicated hardware. Instead, a user needs to connect to a set of login nodes (typically via SSH) and then submit jobs to a local batch system. Any push model for pilot or job submission needs to work through these login nodes. Pull models rely on services running on login nodes or other edge nodes provided by the HPC. The availability and type of edge node varies from HPC to HPC.

HPC compute nodes frequently do not have substantial direct TCP/IP connectivity to the broader internet, instead encouraging or requiring large amounts of data and interprocess communication (IPC) to happen via specialized data fabrics. Typical experiment workflows, however, often send updates regarding job state and progress via some control channel (e.g., outbound connectivity to a RESTful HTTP endpoint, as in the ATLAS PanDA system) to a central coordination system. In the case of the LCFs, which operate at the most extreme scales, outbound network connectivity is completely restricted, meaning that other mechanisms must be used to communicate between jobs and the centralized workflow coordination system. To overcome these architectural differences, significant effort has been invested to use HPC file systems as an IPC medium. 

For US HPCs, data transfers are facilitated through Globus Online~\cite{bib:globus}, which is broadly supported by many HPC and Supercomputer centers as a standard mechanism for moving data into and out of these facilities. At many institutions, notably LCFs, Globus is the only supported method for wide-area network (WAN) data transfers.

Another area where mapping our grid computing to HPC can be problematic is job size. The larger HPC (especially the LCF) are designed for extremely large footprint calculations that span many nodes (up to thousands of nodes). As such, system design of such HPC focuses on very fast interconnects between nodes and the batch scheduling very often favors such large many-node jobs. In comparison, LHC science is based on collision events. We have a lot of these events, but they are independent of each other and can be processed independently. Therefore our computations are broken down into multiple smaller jobs that run on a single node, traditionally even a single core. In recent years we have developed multi-threaded frameworks to make use of multiple cores in a single job, but for efficiency reasons we operate with only a few cores used by each process. Work is ongoing to improve multi-threading efficiency, but at the same time the number of cores per node also seem to be increasing (a Perlmutter CPU-only node has 256 cores for instance), so it's unclear if we will ever get to a point where we can run a single job taking up a whole node without an unacceptable low cpu efficiency.

What we are doing to work around this at HPC is to provision whole nodes (depending on the exact batch scheduler policies sometimes many whole nodes) and then fill these resources with many of our smaller jobs, with each job only taking up a few cores on a single node at most. Packing many of our jobs of exactly the same length into one batch job would be ideal, but isn't completely possible due to variations in event processing times, even within the same workflow. It also makes integration more complicated since it imposes restrictions on how we use HPC resources (compared to grid resources, that are expected to be able to run a mix of workflows). Alternatively, we can accept inefficiencies in filling the batch resources. What that means in practice is that on a many node batch job, toward the end of the wall time, we will gradually ramp down using the resources (within a node and among nodes) until eventually all of our jobs complete and we terminate the batch job. 

Given that scheduler policies are different at every HPC site, some care is needed to organize workflows such that jobs run with good efficiency while maximizing throughput. When integrating a new resource, we must consider the minimum and maximum nodes per job, maximum number of jobs in queue, CPU cores per node and maximum job run time (among other attributes) to come up with a reasonably well-shaped job for every site.

\subsubsection{ATLAS}

Today, ATLAS integrates NERSC Cori, NERSC Perlmutter and TACC Frontera via the Harvester~\cite{bib:harvester} edge service that runs locally to the HPC. Harvester facilitates communication between the PanDA ~\cite{bib:panda} workflow manager and the HPC site by monitoring job states in both PanDA and local batch queues, initiating data transfers as needed, and managing the overall job life cycle. 

All storage resources in ATLAS must be registered in the Rucio ~\cite{bib:rucio} data management system, and any data needed by workloads running on HPC sites must come from a known Rucio Storage Endpoint (RSE). Likewise, any data generated by workloads must be registered in Rucio for downstream consumption. As HPC sites today do not support any form of custodial storage, a proxy RSE has been established at the Brookhaven National Laboratory Tier 1 facility, for the sole purpose of facilitating data transfer between the Rucio data federation and US HPC sites. 

\begin{figure}
\centering
\includegraphics[width=\textwidth]{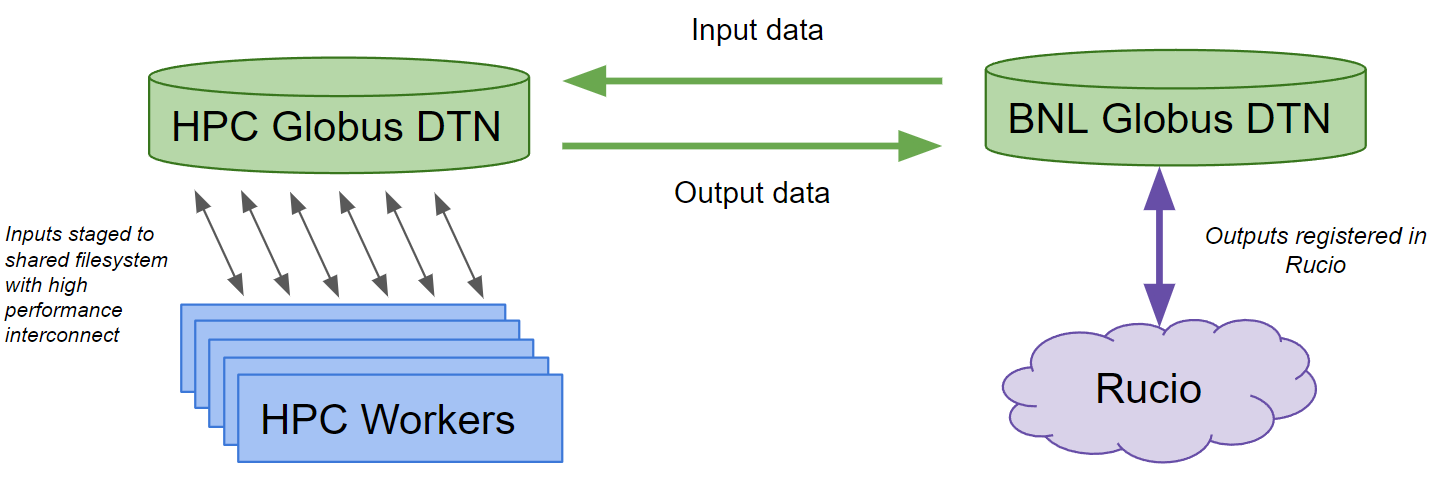}
\caption{Diagram of data flow between HPC and Rucio for ATLAS jobs}
\label{fig:bnl-nersc} % Give a unique label
\end{figure}

For a given workflow, PanDA will assign jobs to the HPC site and data to the Rucio Storage Endpoint at the BNL T1. Input data will be transferred as necessary to the RSE at BNL, jobs will become 'activated' at the HPC site, and the Harvester software will initiate a Globus transfer request between the Brookhaven proxy server and the endpoint at the HPC site. After the input data transfer has completed, Harvester will submit workloads to the HPC scheduler, monitor the jobs until completion, and then start a Globus transfer in the reverse direction to stage out any output files for registration in the Rucio federation. A diagram of this workflow can be seen in Figure \ref{fig:bnl-nersc}.

\subsubsection{CMS}

CMS has chosen to integrate NERSC, XSEDE HPC and TACC Frontera into our normal GlideinWMS/HTCondor based Submission Infrastructure~\cite{bib:glideinwms}~\cite{bib:htcondor}~\cite{bib:htcondor2} by making them look like a grid site. As such, we can target workflows at HPC sites in exactly the same way we target them at grid sites, by including the HPC site in the workflows allowed site list during assignment.

The underlying technical integration works different to a grid site, since HPCs don't provide a local Grid CE to interface their batch systems. For US based HPC we instead work through the HEPCloud portal at FNAL. Within HEPCloud we integrate HPC in two ways. First is direct pilot submission via Bosco~\cite{bib:bosco}, which we use for NERSC. For XSEDE HPC and Frontera we rely on Hosted CE~\cite{bib:htcondorce} run by OSG, which internally also use pilot submission via Bosco. Otherwise it's a normal job pressure based resource provisioning model, jobs targeted at HPC are making it into our workflow management system and this triggers provisioning of HPC resources through the submission of pilots. Eventually the pilots run and execute the CMS jobs. Due to queue times in HPC batch system, often the pilots have to stay pending for some time before they run and this means the CMS jobs that originally triggered them will have run somewhere else in the mean time. But as long as we can maintain continuous job pressure for HPC resources, this does not matter, the pilots will just execute other CMS jobs targeted at HPC. This doesn't necessarily mean utilizing HPC resources will require extra effort, but it does mean that in this model a good HPC utilization depends on a large enough fraction of the workflow mix that is in the system at any given time being able to run on HPC. If the workflow mix changes in a way that this isn't the case, HPC utilization will drop. We keep an eye out for this to happen, but during this year only observed it a few times, always being a temporary situation and overall having a small effect.

We avoided the data transfer problem so far by relying on direct streaming of job input data from remote CMS storage via AAA. Depending on the exact HPC we run at and the scale of operations there, this can be more or less successful. We've had very good success at XSEDE HPC, but our allocations there are smaller with consequently smaller amount of peak resource utilization and the XSEDE HPC also have good outbound network connectivity from the worker nodes. At larger scales of operation or with problematic outbound network connectivity from the worker nodes, job efficiency can suffer or at worst jobs can fail. In the absolute worst case scenario the input data streaming to CMS jobs can cause problems for the whole HPC sites external connectivity. We had the latter problems at TACC Frontera, where very large allocations and consequentially very large peak utilization resulted in network loads that triggered instabilities in the site firewall. We had to limit peak utilization of the resource to find a stable working point. 
At NERSC we initially had severe problems utilizing streaming of input data to job, due to technical limitations with the outbound network connectivity from worker nodes. After some efforts from NERSC this limitation was removed~\cite{bib:nersc_sdn}, which allowed us to utilize NERSC resources with workflows that needed streaming of input data. In later years NERSC also provided CMS with a large scratch quota, enough to store pileup data for the most current MC campaign.

Notwithstanding these issues, we've managed to find a stable operating point for all these HPCs and our chosen integration method has worked very well for us. It does rely on the HPC worker nodes having outbound connectivity to the internet. This is something that only works at the User Facility HPC, and not at the LCF.

Integration of LCF into our computing infrastructure is an active R\&D area. We have used a prototype integration into the HEPCloud portal to produce HL-LHC MC Samples for Snowmass, but the stability and performance of the prototype means that more R\&D is needed before we can consider expanding use of the LCF.

\subsection{Further HPC R\&D}
\subsubsection{Planning for future workflows}

Both ATLAS and CMS started targeting workflows at HPC that are "easy" to run there. This has evolved/expanded somewhat over time and what is "easy" can also be different for us. But the general approach is still the same, we both limit selection of workflows we target at HPC.

So far that approach has not caused problems, we both have enough work to run on the HPC resources available to us. Looking forward to a possible HL-LHC future where a larger fraction of our total resources potentially come from HPC, the question is whether this approach will stay feasible. We should keep working on removing limitations and enabling a larger fraction of our workflows to be able to run on HPC.

What complicates the matter is that the LCF, which are the HPC with the biggest growth potential for us (both due to their size and lack of current use) are also the ones with the most difficult architectures (GPU) to make use of and the ones with the most restrictions in terms of integration into our overall computing infrastructure.

Further R\&D to enable running as many of our workflows as possible on LCF is therefore a top priority for the future.

\subsubsection{Reducing Operational Effort via Service Platforms} \label{service-platforms}
HPC centers are increasingly prototyping edge service platforms that service their user communities with Kubernetes-based clusters, for the purpose of facilitating long-lived services that can batch work to an adjacent HPC. Examples of this include Spin\cite{bib:nersc-spin} at NERSC, Slate\cite{bib:olcf-slate} at OLCF, and PetrelKube\cite{bib:alcf-petrelkube} at ALCF. There is an opportunity to collaborate with these institutions and influence the design of these systems while the technology is nascent. Based on our usage of HPCs today, we would benefit greatly from the ability to create long-lived services that run in a service platform external to the HPC, which can act as an intermediary for facilitating workflows and data transfer to and from the site. This can be done in one of two ways, based on our current experiences with these tools: 
\begin{itemize}
    \item Interaction with the HPC via commandline tools and mount points within the container, meaning that the Kubernetes platform will need to have some means to make available batch system and data transfer tools to the user. This is broadly the approach taken by the OLCF Slate platform, which provides wrappers for 'sbatch', 'squeue', etc. 
    \item Interaction with the HPC via RESTful API, such as the "Superfacility API" being developed by NERSC \cite{bib:nersc-superfacility}, which would allow services to interact with NERSC supercomputers, data transfer services, etc via a common API. ATLAS and CMS would need to invest in corresponding integration for this API, e.g. Harvester plugins (ATLAS) or GlideinWMS plugins (CMS). 
\end{itemize}

It is important to note that a ubiquitous standard workflow for accessing HPC resources such as compute and storage via service platform has not yet emerged and it is not yet clear that there will be one, so some caution must be taken when considering moving integration efforts there.

\subsubsection{Rucio \& Globus Online} \label{hpc-rucio-globus}

HPC sites frequently have substantial storage resources available, but offer that storage in a way that is somewhat incompatible with our usual expectations. The Rucio software team has recently implemented a mode to manage Globus Online endpoints and treat them as a Rucio Storage Elements (RSEs). If Rucio can seamlessly treat HPC storage endpoints as RSEs, it may be possible to request large storage commitments at HPC sites as a part of the allocation request process, integrate them into our data management fabric as standard RSEs and use them as a storage nucleus for HPC work. 

Even so, we would still be limited to transferring between Globus Online endpoints and XRootD endpoints respectively. In order to have a complete transfer mesh and allow transfer from any storage to any other storage we would have to rely on some storage supporting both protocols (and being able to act as an intermediary).

In any case, if we plan to use large amounts of HPC for compute we will need to provide a corresponding amount of storage somewhere. It would be ideal to build this into an HPC allocation, otherwise we will have to shore up the difference somewhere else in our computing fabric. 

\subsubsection{Hosted CEs}
For some time, the Open Science Grid (OSG) and Partnership to Advanced Throughput Computing (PATh) have been developing a Software-as-a-Service to connect batch resources, including HPCs, into scientific computing fabrics. This Hosted Compute Element (also referred to as a Hosted Compute Entrypoint or Hosted CE) has seen production use for the CMS experiment on several HPC, including TACC Frontera/Stampede2, PSC Bridges-2, SDSC Expanse and Purdue Anvil. One advantage of the Hosted CE is that maintenance and operation of the Grid-specific middle-ware is lifted into a central service infrastructure, maintained by the developers of the Hosted CE application. Site operators, such as those who operate HPCs, need only provide a standard secure shell (SSH) interface, which is already ubiquitous at all computing sites and therefore adds almost no overhead to facilitate ATLAS and CMS workloads. This mode of operation is advantageous to the experiments as well, as it moves some of the work needed to adapt workflows to a third party provider.

\subsection{HPC Cost}
% No direct comparisons - probably.
% We need to do the work collecting information now, but we should write this last. It is going to be political
% Direct vs Indirect costs
% Direct
%    - hardware
%    - people
% Indirect
%    - other parts of the facility impacted by HPC/Cloud

% We also need to do costs of R&D and longterm operations.

% We should do costs for today - TCO for Cloud and HPC.
% Let's also extrapolate various scenarios and their cost impacts. This will be partially informed by the workshop.

As HPC allocations in the United States require competitive proposals in order to receive allocation time, there's a non-trivial amount of effort spent in writing proposals with strong physics use cases, reporting back to the facility with results produced by time spent on resources at the HPC and so on. 

While HPCs have no direct hardware cost for ATLAS and CMS, it is worth considering the costs of ancillary services that must be maintained, in terms of capital expenditures as well as employee time. 

\subsubsection{ATLAS}
Operationally, ATLAS is currently maintaining Harvester instances at both NERSC and TACC with 0.55 FTE in the HPC operations area of the US ATLAS operations program, which includes integration and ongoing maintenance including keeping the Globus endpoints activated, troubleshooting job failures, keeping the ATLAS Pilot and Harvester software up to date, and so on. 

The ATLAS Tier 1 at Brookhaven National Lab invested in a dedicated Globus Data Transfer Node (DTN) to facilitate transfers between our principle HPC targets at TACC and NERSC. As ATLAS HPC usage continues, this hardware will eventually age-out and need replacement, or BNL may require additional hardware to meet the demands of the workloads as our reliance on HPC sites for a nontrivial fraction of ATLAS workloads grows over time. 

In addition to hardware, some considerable fraction of person time is spent on maintaining the hardware, storage system and other software services running on this machine. This may need increased effort if ATLAS chooses to rely more heavily on HPCs for future workflows.

\subsubsection{CMS}
CMS is accessing US HPC through the HEPCloud portal at FNAL. The HEPCloud portal is not just supporting CMS, but also other FNAL experiments. The US CMS Operations Program is currently contributing 0.5FTE to HEPCloud Operations, but this only covers running CMS workflows at the User Facility HPC, not any efforts targeted at LCF. There is additional R\&D efforts for the LCF, once this R\&D pays dividend and LCF use becomes more common, there will be a need to increase HEPCloud Operations effort contributions (we are already considering this for FY23).

Apart from effort costs, there are also indirect hardware costs. The current integration of the User Facility HPC, with streaming of input data to HPC jobs and direct stageout from HPC jobs to FNAL storage, contributes to a higher load on the networks connecting US CMS sites to HPC and also leads to a higher load on storage at US CMS sites. FNAL as the target site for stageout from HPC and also the site with the largest storage pool (and therefore the highest chance to host input data needed by HPC jobs) is mostly affected by this. So far these indirect efforts have not lead to any problems and the existing FNAL infrastructure could handle the load spikes from HPC. In a possible (HL-LHC) future with potentially much larger HPC contributions, we would likely have to take this into account for FNAL hardware provisioning decisions.

Future LCF operations are even more uncertain, since there is still ongoing R\&D in this area and the final operations model hasn't been decided yet. No matter what this will look like in detail, it is clear that to utilize a large number of cores at an LCF, we somehow have to bring input data (for jobs that require input data) to these cores and then also move the results of the computations at the LCF back to CMS storage. Depending on the scale of LCF operations this will have an impact on network interconnects and storage systems at CMS sites.

\subsubsection{Common costs and projections} 
There are also a number of ongoing FTE expenses in terms of effort expended by both ATLAS and CMS staff to do initial integration and ongoing operations of their respective computing frameworks with HPC sites. 

For the initial integration we are in a good space, having learnt many lessons over the last few years integrating various different US HPC. US HPC now usually support some container system and CVMFSExec works well on newer Linux kernels (providing access to our software then becomes a question of the availability of an edge squid proxy). The current generation of HPC, at least in the US, seem to be dominated by the \textit{x86} CPU architecture, which simplifies having compatible software for it. Even if that should change (ARM being deployed at US HPC for instance), supporting different CPU architectures in principle only requires recompiling our software stack and then a physics validation of the new architecture. Both of which require work, but the process is at least straight-forward. CMS just went through this for Power (for the Marconi 100 HPC in Italy, which is the same system architecture as OLCF Summit). This only covers different CPU architectures though, it doesn't address how we can add support for GPU architectures that are being deployed at US HPC, especially the LCF. That is a much more complicated problem, which requires internal changes to our event processing frameworks and these efforts are outside the scope of this report.

If we end up with larger storage quotas at LCF, we can also foresee effort needed to operate these storage quotas as part of the overall ATLAS/CMS data management systems. This will require additional effort compared to the same storage quotas at our own sites due to the special handling needed for Globus Online transfers.

Some of the effort projections are subject to change based on the successful (or unsuccessful) outcomes of R\&D projects to move more grid-like storage closer to the HPC facilities, or to more closely integrate Globus with the existing Rucio Data Federation technology used by both ATLAS and CMS.

\subsection{International HPC efforts} \label{hpc-international}

% e.g. Vega, Karolina in ATLAS
% BSC/Marenostrum
% Cineca-Prace

\begin{figure}
\centering
\includegraphics[width=\textwidth]{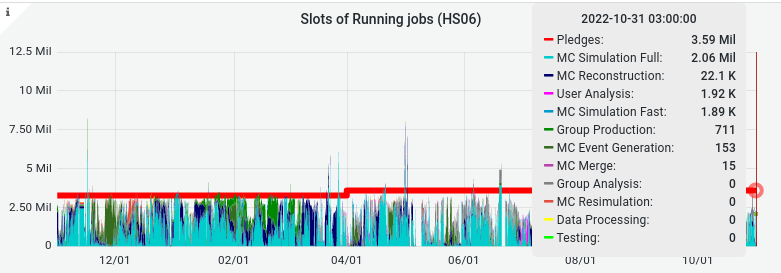}
\caption{Running jobs on EuroHPC Vega over the last year, compared to the total ATLAS pledge}
\label{fig:vega-peldge} % Give a unique label
\end{figure}

The EuroHPC initiative started in 2018 with the goal of building world-class HPC infrastructure for Europe. The collaboration consists of 31 countries, mostly European Union members, and is governed by a Joint Undertaking board. In 2018 through 2021, the EuroHPC Joint Undertaking invested 1B€ in supercomputing, quantum computing, service and data infrastructure ecosystems. In the first phase of the initiative, five new Petascale (5-20 PetaFLOPS each) and three pre-Exascale HPCs (250-350 PetaFLOPS each) were funded and hosted by various member states. Seven of these systems are in pre-production or production and have made significant contributions to ATLAS and CMS computing already.

Systems of note include the Vega HPC hosted by the Institute of Information Science in Maribor Slovenia (IZUM). Vega has about 1,000 nodes in the CPU partition and 60 nodes in the GPU partition. LHC experiments had early-stage input in the design phase and accommodated some of ATLAS requirements, making the extensive usage of Vega possible. On average, Vega has been delivering 140k CPU cores since April 2021 to the ATLAS computing effort. This is roughly a fifth of the overall ATLAS Grid CPU resources during that period.

For several years, ATLAS has opportunistically made use of the computing resources at the IT4Innovations (IT4I) National Supercomputing Center in Ostrava, Czech Republic. One of the newly commissioned EuroHPC machines, Karolina, is an AMD HPC with 720 CPU-only nodes, hosted by IT4I. In the last quarter of 2021, this machine provided significant cycles to the ATLAS collaboration, with additional usage pending project renewal. 

EuroHPC supercomputers, like US HPCs, are driven by an ever-greater need for floating point performance. While many of the Petascale machines procured by EuroHPC offer significant CPU-only resources, the larger pre-Exascale machines are dominated by GPU accelerators. For instance, while less than 10\% of the nodes in the Vega HPC are GPU-enabled, GPU nodes make up over 60\% of the LUMI supercomputer (the premiere EuroHPC supercomputer, \#3 on the Top500 in May 2022).

\subsection{Desirable HPC facility features and policies}

CMS and ATLAS have fundamentally similar computing models, sending approximately the same types of workloads to the same sets of HPC providers. Correspondingly, it tends to be the case that the same HPC challenges faced by one collaboration, such as workflow scheduling and data movement, is also faced by the other collaboration. It would be very desirable for facilities to consider the needs of large, multi-institutional experiments such as ATLAS and CMS, but also other (perhaps non-HEP) collaborations or those who approach (or even exceed) the scale of LHC computing. This goes beyond just submitting workloads, as centers will need to also consider the data needs of collaborations both in terms of capacity and wide-area throughput. Finally, in order to integrate into large distributed computing fabrics, HPC sites need to consider evolutionary steps for their security models that will facilitate automation of HPC resources as part of computing in larger collaborations.

\subsubsection{Batch Policies}

As previously mentioned, the focus of larger HPC (and especially the LCF) on only the largest computing problems that require many nodes (up to thousands) all interconnected within the same single computation doesn't match particularly well to the LHC science case of studying many independent small collision events. This causes inefficiencies in how we map the size of our computational units (i.e. single jobs with defined start and end time) to the size of the resources provisioned with a single HPC batch job.

Allowing single node HPC batch jobs would help reduce these inefficiencies. Even with single node HPC batch jobs allowed though, the batch scheduling policies would need to not penalize them indirectly. At NERSC for instance single node batch jobs are allowed, but the way the batch scheduler is configured one cannot run them very frequently (assuming a busy system), so in practice one is forced to submit many node batch jobs (otherwise it is practically impossible to use up the hours in a larger allocation).

If this isn't possible and we have to provision HPC resources as many node batch jobs, we would at least require them to be provisioned with longer wall times. Allowing longer wall times on HPC batch jobs helps reduce ramp down inefficiencies introduced by varying wall times of our jobs running within the HPC batch job.

\subsubsection{Security Model Considerations} \label{mfa}

HPC facility security models are increasingly moving toward a security posture, where multi-factor authentication (MFA) is the only acceptable method of user authentication to a site. MFA improves traditional authentication schemes by requiring additional pieces of identifying information from a user trying to access the system, which are generally expected to be something that the user knows (e.g. a password) and something that the user possesses (e.g. a one-time token generated by a mobile phone application such as Duo Mobile or Google Authenticator). 

%In practice, ATLAS and CMS have thousands of physicists submitting work to their respective global job management systems, which ultimately get multiplexed through a small number of user accounts (often 1 user) at each HPC facility. 

While this model works well for individual (or small groups of) researchers submitting tasks to an HPC by hand via interactive shell, the process by which workloads are submitted to computing resources for ATLAS and CMS would ideally be entirely automated, modulo initial setup. Sites with restrictive MFA policies that require a human interaction component make automation extremely difficult. As such there is a general, demonstrable need to broaden acceptable authentication factors in HPC site policies to include some additional authentication element that does not preclude automation, while still meeting the security needs of these facilities. In practice, this has been implemented at some amenable facilities by adding a limited set of source IP addresses that represent the Hosted CE service to an allow list (either at the firewall or SSH daemon level) as an inherence-based authentication factor, combined with key-based authentication as an additional factor.

The more or less amenable divide between HPC seems to fall along the existing divide between LCF vs. User Facility. This is another reason that makes integration of the LCF into the ATLAS and CMS computing infrastructure more difficult.

\subsubsection{Storage} \label{hpc-storage}

% we have a paragraph or two on this before, we should consolidate this somehow

When it comes to storage use at HPCs, there are two different aspects to consider. First is job scratch space for actively running jobs (runtime scripts, temporary files generated by running processing, etc). At grid sites this is usually local storage directly attached to the worker node (nowadays flash based). Some HPCs (usually in the User Facility category) also have node-local, flash-based storage available. This is very useful for HEP workflows that often deal with many small files.

At the larger HPCs, and especially at the LCF, often the only option available for this is space on a large shared file system. These shared file systems usually aren't well suited for the large number of concurrent meta-data operations from the many parallel running HEP jobs. There are workarounds available for this (basically bind mounting a file system on the node onto a file on the shared file system), but they increase the integration complexity.

% do we need references to NodeCache and the corresponding singularity feature?

A second use of storage for our workflows is for job input and output data, i.e. collision events (both real data and MC). How much of such storage is needed is highly workflow dependent. For instance, if one only runs MC simulation workflows, starting with event generation, then there is no input data. A data reconstruction workflow on the other hand would need to have access to RAW data.

Availability of outbound internet access from worker nodes to experiment owned storage at other sites (either directly or through a proxy edge service) allows integration models where input data reads and output data writes from jobs can bypass the local storage available at the HPC, at the cost of larger loads on networks and on the storage at our own sites.

At the LCF, where outbound internet access from the worker nodes is not available, we would need larger storage quotas to host job input data for processing to allow more flexibility in the types of workflows we can run there. Indirect access to external, experiment-owned storage through proxy edge services might be a possibility for the future, but even if this could be deployed at the functional level, it's unclear if such a system would ever be feasible to operate at scale, given the potential very large amount of compute resources available at the LCF.

% CMS uses minimal storage in CFS. They keep the most common pileup library in scratch. Somewhere between 500-700 TB. Increases job efficiency.
% 
% At the LCFs: CMS asked for 200TB, got 100.
% Dirk notes it may not be performant enough - to be discussed, maybe good for the doc. 
% For 

\subsubsection{XRootD Services for LHC Collaborations} \label{xrootd-service}
While the largest HPC sites generally provide dedicated Data Transfer Nodes (DTNs), typically only the proprietary Globus Online service is supported for large-scale transfers. While Globus is appropriate for many of the users that HPC facilities service, LHC experiments instead rely on the XRootD protocol and SciToken-based authentication for large scale transfer of data between and within WLCG computing sites. XRootD is ubiquitous in the HEP computing community with two well-established implementations (XRootD and dCache XRootD) used in production for well over a decade. Both ATLAS and CMS have recently been working together with the NERSC facility to provide an XRootD-based DTN for use by both experiments. If successful, this approach will significantly simplify integrating NERSC into the respective data federations for CMS and ATLAS, reducing unnecessary proxy transfers via the WAN while making the facility appear more similar to other WLCG sites. 

% Let's definitely talk about shared data and submit infrastructure. ie., what do we want to get out of this? I think we want the high-level folks to strongly recommend to HPC sites that they support data/compute gateways that would facilitate ATLAS, CMS, and other workloads. 

% Perhaps also make recommendations about how security models ought to evolve to include automation.

\section{Cloud Computing}
\subsection{Landscape of Cloud}
% Probably the big 3: Azure, Google, AWS

\subsubsection{The Origins}
When discussing the usage of Cloud Computing, and unless we explicitly state that we are talking about a particular case, we are referring to the major commercial cloud providers like Amazon Web Services, Google Cloud Platform, Microsoft Azure, Oracle Cloud and so on. 

Commercial clouds started out by leasing the data center spare capacity, which was not used by their core business activity. For example Amazon built data centers for their online sales, which fluctuate depending on the time of the year and particular events like Valentine's day or Black Friday. Google requires the compute power and storage for their Search, YouTube and Gmail. The load will depend on the time of the day, during which users have a higher online presence. Therefore data centers are scaled to cope with the peak demand and leave free cycles during other periods. Their business models evolved and cloud computing became one of their core activities. They offer some of their internal technologies to the clients and build on economies of scale to operate gigantic data centers for their own and their clients' computing needs. 

\subsubsection{Global Locations}
Clouds are geographically distributed and structured in Regions and Availability Zones, to place the resources closer to the end users. Each Region is a separate geographic area, where a few data centers are clustered. For example Amazon has four Regions spread across the USA (Oregon, California, Ohio, Virginia), six Regions spread across Europe (Ireland, UK, France, Germany, Italy, Norway) and more in other continents. Each of these Regions contains a few Availability Zones, e.g. three to six in USA and Europe. An Availability Zone is one isolated data center with redundant power, network and connectivity within the Region. 

The capacity of Regions and Availability Zones are confidential information and not disclosed to the public, but are expected to be much larger than any WLCG computing center. However even (or in particular!) the largest cloud providers carefully optimize their capacity and try to avoid having too many idle resources. The usual perception of "infinite" computing power and bandwidth needs to be taken with a grain of salt. 

When integrating commercial cloud resources with the experiments, the choice of one or multiple Regions requires some attention. Usual reasons for the choice of the Region or multiple Regions are:
\begin{itemize}
    \item Size of the Regions and the cloud deployment: we need to select a large enough Region for our needs, or might even want to break it across multiple Regions. Cloud-capacity teams usually advise on suitable Regions, while not disclosing too many details.
    \item Availability of the required resources: some resources, e.g. the latest processors or GPUs, are not available in all Regions. The availability of these usually gives an indication of the Region's size.
    \item Geographical location and availability of network peering: we want the Region to be geographically close to CERN or a particular WLCG site. Some Regions also have dedicated interconnects with National Laboratories through ESNet ~\cite{bib:esnet} or Google Zurich has been peered with CERN. This choice affects the latency and bandwidth when transferring large amounts of data.
    \item Other reasons, such as the Carbon emissions for the Region: some cloud providers advertise each Region's CO2 emissions~\cite{bib:gcp-carbon} and customers can choose environmentally friendly locations. 
\end{itemize} 

The cloud providers operate global networks to provide high connectivity between Regions. The more distant the Regions, the more expensive the network is, and the cost of the data transfers will eventually be passed to the customer.

\subsubsection{Service Levels}
Commercial cloud providers offer services across different levels:

\begin{itemize}
    \item \textit{Infrastructure as a Service} (IaaS): provisioning of low level services like servers, storage and network.
    \item \textit{Platform as a Service} (PaaS): provisioning of development tools, database services, analytics services and others.
    \item \textit{Software as a Service} (SaaS): hosted applications spanning anything from email, calenders and document management to various business applications.
\end{itemize}  

In recent years, intermediate service levels are emerging, most notably the one coined \textit{Container as a Service} (CaaS), where the users can run their containerized applications on hosted Kubernetes ~\cite{bib:kubernetes} platforms or serverless alternatives, where the infrastructure is hidden from the customer.

LHC experiments will typically use \textit{Infrastructure as a Service} and \textit{Container as a Service} levels for integrating the cloud with their distributed computing frameworks or their on-premise infrastructure. The next sections will describe different integration possibilities in more detail.

\subsubsection{Cloud benefits}

Considering all of the above cloud features, some of the important benefits to exploit are:
\begin{itemize}
    \item Elasticity: it is possible to scale out large deployments for a short period of time, thereby reducing the \textit{Time to Science} and renting the resources only for the time they are being used.
    \item Availability of heterogeneous resources: some resources, which are not widely available at grid sites, can be evaluated on the cloud. This is the case for ARM processors, GPUs or FPGAs.
    \item Reliability and ease of use: cloud providers benefit from economies of scale and running the same services for many customers. They invest significantly in reliability and ease of use of their infrastructure. 
\end{itemize}  

We will give examples for all of these advantages in later sections. Of course the advantages come at a monetary cost and raise some concerns within the community, which we will discuss in section ~\ref{Concerns}.

\subsubsection{Special Clouds}\label{special-cloud}
It is worth mentioning that other newcomer clouds are getting attention by differentiating themselves through unique characteristics, with one example gaining attention in the WLCG being Lancium \cite{bib:lancium}. We will discuss the Lancium features and integration shortly in section ~\ref{cloud-cms}.

\subsection{Cloud Projects in the Experiments}

ATLAS has had multiple interactions with commercial clouds in the past. ATLAS pioneered the usage of cloud computing as a flagship project ~\cite{bib:atlas-helixnebula} during the first stages of the European HelixNebula~\cite{bib:helixnebula} initiative between 2012 and 2014. ATLAS also ran short term demonstrators on Amazon and Google ~\cite{bib:atlas-cloud} during Run 2, showing promising results and scale. In more recent years, longstanding projects have been started with Google through a series of US ATLAS funded projects (2018-2022) and Amazon through a California State University Fresno grant (2020-2022). These projects have established relationships that go beyond just renting out infrastructure. In particular with Google, we have been meeting on a regular basis, exchanging information about technologies and the experiment computing model. Google organized presentations on particular topics of interest or offered a multi-day training on cloud architecture for a group of ATLAS participants. The continuous interaction helped to shape the current integration model and introduced the experiment to industry practices.
The Google project was extended in July 2022 and became a project funded by the whole ATLAS collaboration. Its purpose is to run the equivalent of an ATLAS site on Google Cloud Platform (from now on the ATLAS Google Site) with an average of 10k vCPUs (virtual CPUs, equivalent to a hyper threaded core) and 7PB of storage over 15 months. This project will have a dedicated track to evaluate the Total Cost of Ownership (TCO). TCO results and findings are expected to be published in the second half of 2023 and not ready for this blueprint document. We will therefore provide interim cost estimations, while a more detailed study should be available later in the year.

CMS started interacting with commercial clouds in 2016 through the Fermilab HEPCloud pilot project~\cite{bib:cms-aws}. This pilot project was created to evaluate the merit of the HEPCloud facility concept~\cite{bib:hepcloud}, a portal to an ecosystem of heterogeneous commercial and academic computing resources with a powerful decision engine capable of routing user workflows to on-premises and off-premises resources based on efficiency metrics, cost, workflows requirements, etc. Production  workflows chaining all stages in the simulation were produced and ran on Amazon Web Services and the Google Cloud Platform with positive results. While we do not use this technology at present for production, due to cost-driven reasons, the infrastructure is maintained up to date in case the experiment wants to use it at any time. 

Another cloud resource that has been integrated in the CMS experiment is Lancium~\cite{bib:lancium}. At present, this resource has been validated with a few small production workflows.

Besides this, CMS is currently focusing on dynamic usage of cloud resources in specialized workflows that rely on machine learning inference servers~\cite{bib:cms-sonic}, to peak usage of both CPUs and GPUs for specific periods of time in order to maximize throughput.

\subsection{Cloud Workflows}
\subsubsection{ATLAS}

\paragraph{WLCG cloud site}
Major commercial clouds provide the building blocks for storage and compute, so that users can easily mix and match components to architect powerful and elastic infrastructures.

Thanks to the stable relations with Amazon and Google, PanDA and Rucio have integrated major providers in a cloud-native way, using standard protocols and services (see Section \ref{hpc-integration} for details). Additional services required to run a performing Grid site (CVMFS and Frontier Squid ~\cite{bib:frontier}) can also be deployed on the cloud. Bandwidth between compute and storage inside the cloud is generally sufficient to run the various ATLAS workloads. During proof-of-concept phases at Google we have focused on running \textit{analysis} tasks for individual users and Monte Carlo processing chains (\textit{event generation, simulation, pile up, derivation}) with varying input/output sizes between O(100MB) and O(1GB) per job. 

\begin{figure}
\centering
\includegraphics[width=\textwidth]{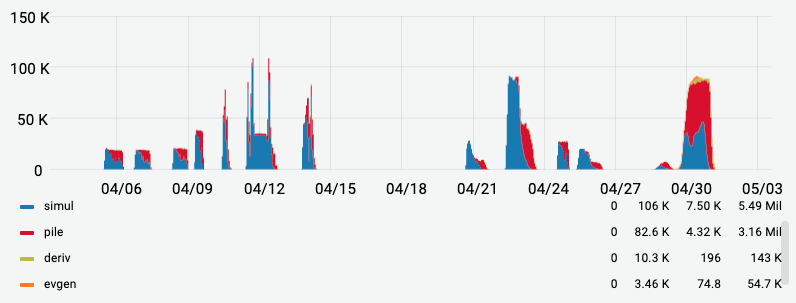}
\caption{Elastic scale out at Google cloud.}
\label{fig:elastic-google} % Give a unique label
\end{figure}

We were able to elastically scale our cloud deployment to up to O(100k) vCPU in a single cluster/Region and executed during periods of around a day the Monte Carlo chains (see Figure ~\ref{fig:elastic-google}).

\begin{figure}
\centering
\includegraphics[width=\textwidth]{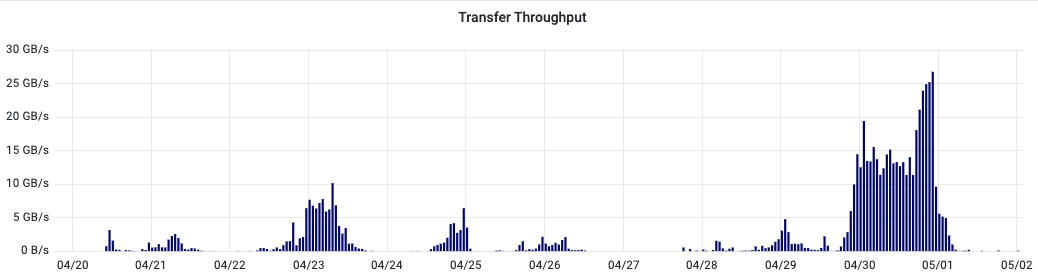}
\caption{Rucio read throughput from data storage to worker nodes.}
\label{fig:rucio-read-throughput} % Give a unique label
\end{figure}

During these scaling tests, the observed read throughput to the workers reached 25GB/s (see Figure ~\ref{fig:rucio-read-throughput}), but without attempting to maximize the throughput. This throughput is comparable to peak read throughput at other large US ATLAS centers. 

\begin{figure}
\centering
\includegraphics[width=\textwidth]{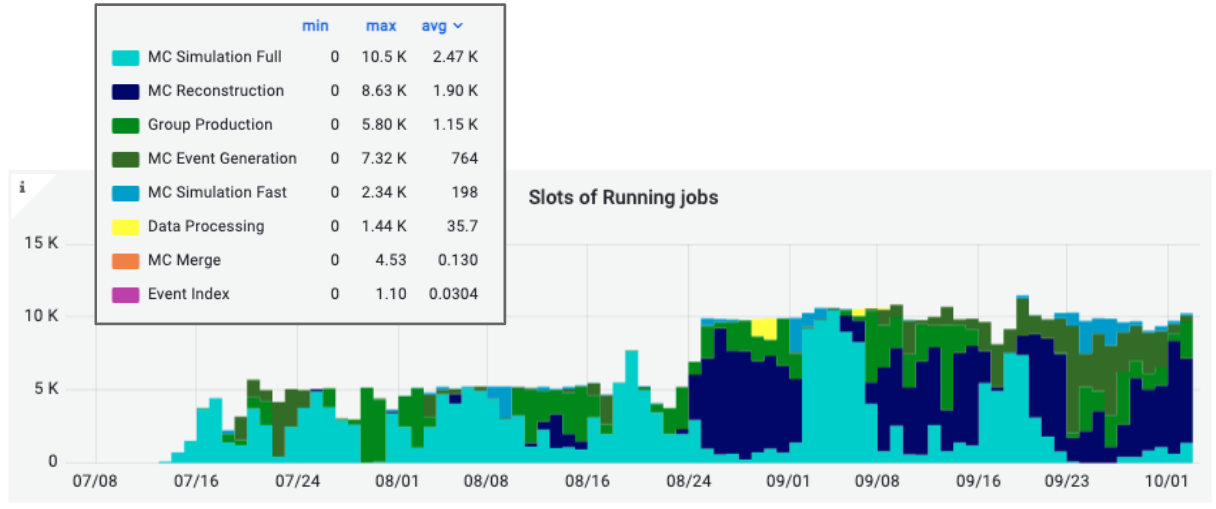}
\caption{Overview of scale and workloads executed at the ATLAS Google Site during the first 90 days of operation between July and October 2022. The cluster started with 5k vCPUs for an initial assessment and was increased to 10k vCPU few weeks later.}
\label{fig:AGP_90days} % Give a unique label
\end{figure}

During the first months of the ATLAS Google Site, we have been operating at a scale comparable to a US Tier 2 center (US Tier 2s are generally much larger than in the rest of the world) and executing any Production workload (see Figure ~\ref{fig:AGP_90days}) assigned by PanDA. The cluster has been running with exceptional stability and we have not observed performance bottlenecks. This means that from a technical point of view it is possible to run a general purpose, Grid-compatible site on the cloud.

% https://monit-grafana.cern.ch/d/000000466/ddm-transfers-historical-data?orgId=17&var-binning=1h&var-groupby=dst_cloud&var-activity=All&var-protocol=All&var-src_tier=All&var-src_country=All&var-src_cloud=All&var-src_site=GOOGLE&var-src_endpoint=All&var-src_token=All&var-dst_tier=All&var-dst_country=All&var-dst_cloud=All&var-dst_site=GOOGLE&var-dst_endpoint=All&var-dst_token=All&var-include=&var-exclude=none&var-exclude_es=All&var-include_es_dst=All&var-include_es_src=All&from=now-30d&to=now

% Talk about elasticity, egress cost and running full chains to optimize this cost
We can however think one step further and optimize the exploitation of cloud resources. Commercial cloud providers stand out positively for their elasticity and negatively for the egress costs (i.e. moving data out of the cloud). For this reason, a possibility is to evaluate different operational models for cloud resources. Rather than keeping the cloud compute resources flat, we could try to schedule full chains to the cloud (as opposed to random steps) and scale the cluster to process these ASAP. This would benefit the egress cost, since the intermediate data products would not leave the cloud. It would also benefit the user experience, since the final data products of the full chain would be obtained much faster. The study of these operational models for the cloud is future work. This is a very similar concept to CMS' \textit{StepChain} model mentioned in Section ~\ref{CMS-HPC-Workflows}.

\begin{figure}
\centering
\includegraphics[scale=0.3]{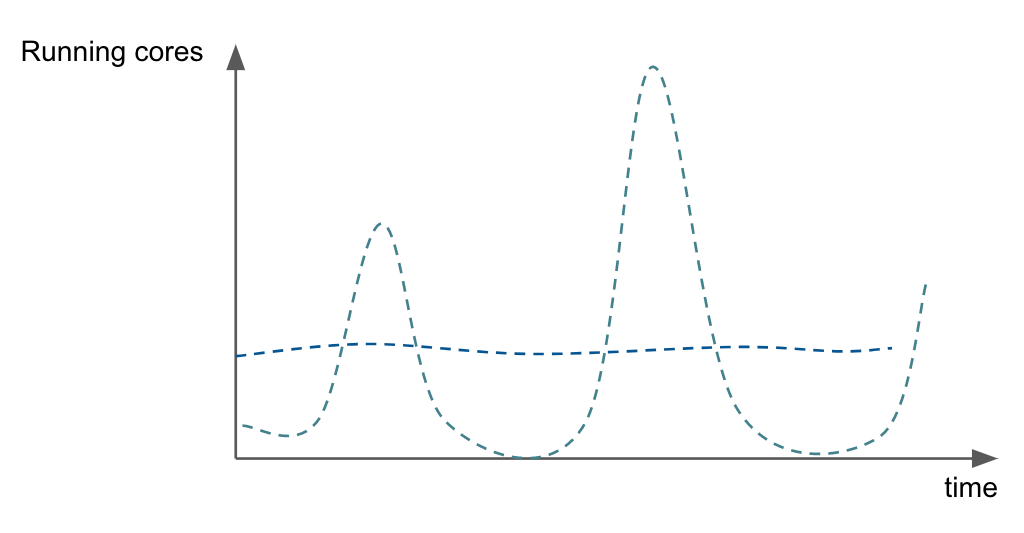}
\caption{Static vs elastic site usage.}
\label{fig:static-elastic} % Give a unique label
\end{figure}

\paragraph{Interactive facilities for user analysis}
The inherent elasticity of commercial cloud resources is also of particular interest for end-user facing facilities. These facilities come in different flavors and include combinations of interactive services (e.g. notebooks, ssh login nodes), local batch systems and parallel computing systems (e.g. Dask, Ray) inspired by other Big Data users and the python ecosystem. 

User facing facilities do not have a consistent load over time, but experience rather spiky load due to users logging in at particular times of the day or producing usage peaks before conferences. It is therefore very difficult to size correctly user analysis facilities and typically they are combined with the general batch processing cluster, where the necessary batch slots need to be evicted for user jobs. 

On the cloud, resources can be auto-scaled dynamically in response to peaks, providing a very good user experience. We implemented an analysis facility prototype on Google Cloud based on JupyterHub (interactive notebooks) ~\cite{bib:jupyter} and Dask (parallel computing framework) ~\cite{bib:dask}. The authentication to the analysis facility goes through the general ATLAS IAM (Identity and Access Management) service and required no additional accounts.

\begin{figure}
\centering
\includegraphics[scale=0.25]{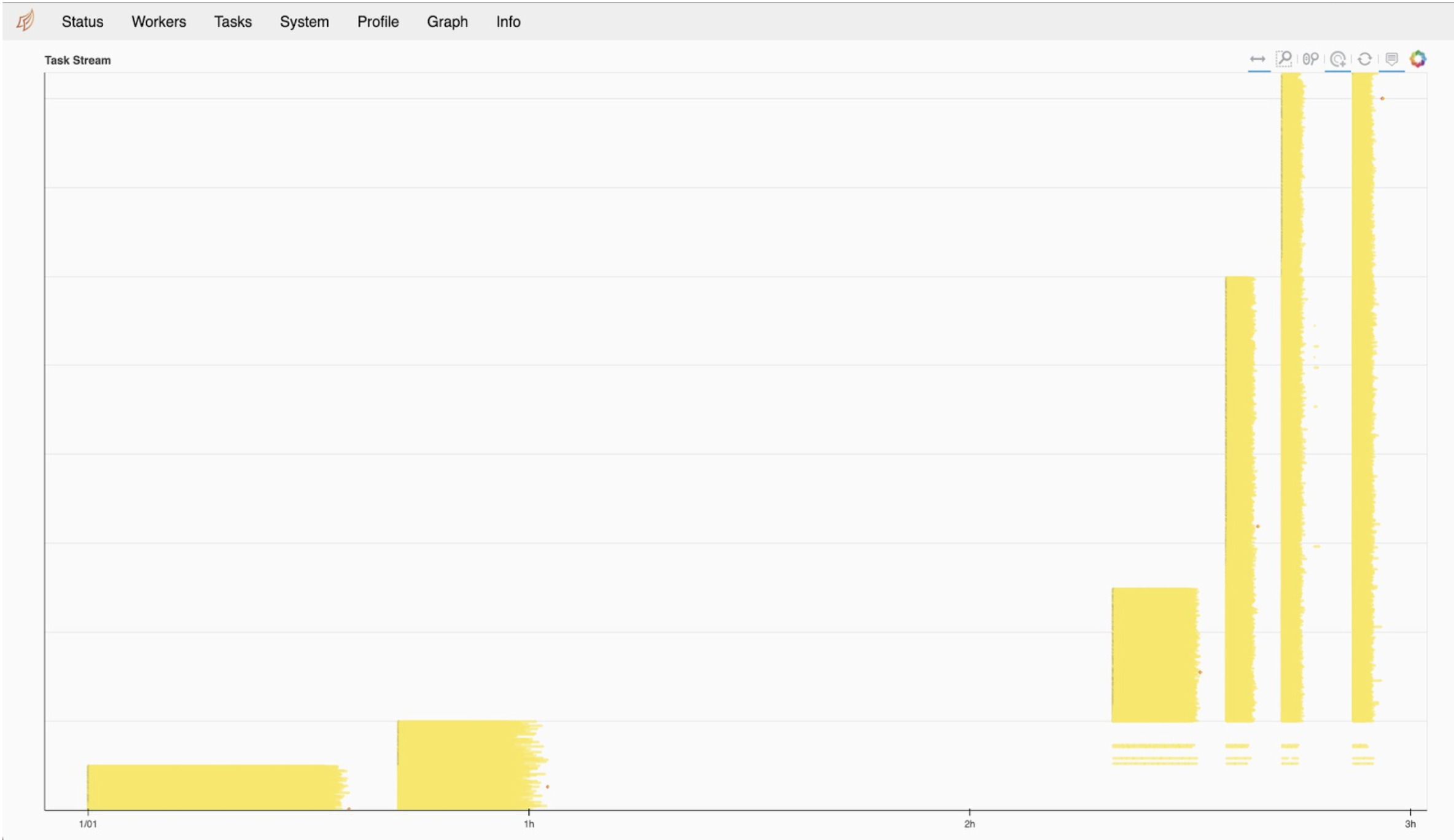}
\caption{Effect of number of parallel workers (y-axis) on execution time (x axis). The same computational task was executed multiple using an increasing number of workers. The duration of the task decreased proportionally to the number of workers, executing within minutes in the last examples.}
\label{fig:dask-scaling} % Give a unique label
\end{figure}

Users can spin up large (we tested up to 4k vCPUs), private Dask clusters and the necessary resources will be available within few minutes. Considering the size of cloud data centers and the resources available, scientists can fully parallelize their computational tasks to get almost immediate results (see Figure ~\ref{fig:dask-scaling}), thus fully empowering them on their scientific endeavours.

It is important to note that these analysis facility systems are still mostly focused towards advanced users and the mainstream analysis is carried out on the Grid or local batch systems.

\begin{figure}
\begin{subfigure}{.9\textwidth}
  \centering
  \includegraphics[scale=0.35]{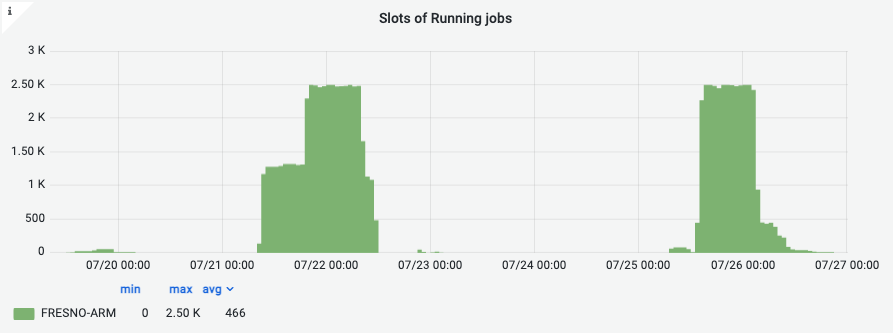}
  \caption{Resources used for the ARM physics validation. The ARM cluster at Amazon (through the Fresno grant) produced 1M events on each bump, using up to 2.5k ARM cores for periods of around a day. The servers shut down automatically as the jobs complete. These resources were sufficient to obtain the official Physics Validation for Athena simulation on \textit{arm64} processors.}
  \label{fig:arm-scaling}
\end{subfigure}%

\begin{subfigure}{.9\textwidth}
  \centering
  \includegraphics[scale=0.35]{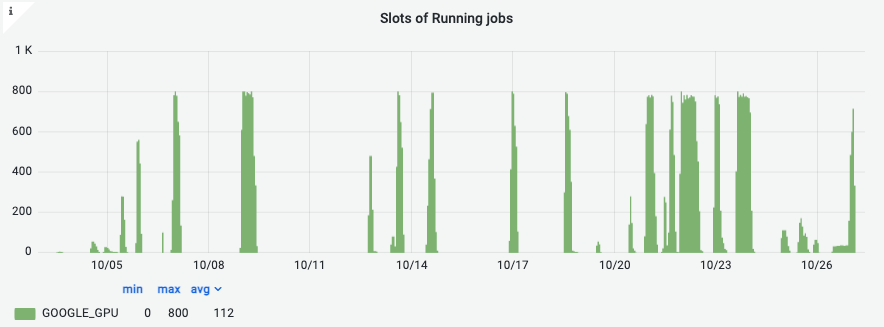}
  \caption{Resources used for studies of the "Off-shell Higgs Measurement Using a Per-Event Likelihood Method" by one of the physics groups at the University of Massachusetts Amherst. Each node with a Nvidia T4 GPU also has 4 cores. The cluster ramps up to 200 GPUs and 800 CPU cores, which is the value seen in the plot. They run iterative bulks of few thousand neural networks that take O(1 day) to complete at this scale.  
  Sidenote: this plot is an example of how the experiment frameworks are built around CPU usage and the GPU statistics are not being collected yet.}
  \label{fig:gpu-scaling}
\end{subfigure}
\caption{Examples of cloud elasticity for resources not largely available at Grid sites. The cost of the resource usage depicted in each subplot is at the order of low-to-mid single-digit thousands of USD, but produced results useful at physics group or experiment level.}
\label{fig:exotic-scaling} % Give a unique label
\end{figure}

\paragraph{Heterogeneous architectures: ARM and GPUs} 
Finally, commercial clouds also enable workloads on heterogeneous architectures, in particular architectures that have not yet been fully adopted by the experiments and are scarce on the Grid. Clouds are a good integration field for GPU clusters and ARM clusters (see Figure ~\ref{fig:exotic-scaling}):
\begin{itemize}
    \item ARM processors are becoming an increasingly attractive alternative to conventional \textit{x86} CPUs due to their low power consumption. For this reason they are present at various HPCs, some Grid sites have expressed their interest and chip makers are integrating ARM and GPU processors in the same chip (e.g. Nvidia's Grace Hopper). The Athena team is starting to build and validate their software for \textit{arm64} processors. It proofed difficult to find on-premise ARM resources, in particular to run a medium scale validation. Instead we set up a PanDA batch queue backed by ARM nodes in Amazon cloud and executed ATLAS first simulation task on \textit{arm64} there. Since then, a full physics validation  was obtained by comparing 1M simulated events on \textit{arm64}  and \textit{x86} processors (see Figure ~\ref{fig:arm-scaling}). 
    \item We provided the possibility to run interactive, GPU-enabled Jupyter notebooks for the development and testing of ML/AI applications. For users that needed to run multiple training iterations, we provided a PanDA batch queue that auto-scales up to 200 GPUs (our current quota) when there are pending jobs. The queue is actively being used by a collaborating ATLAS physics group in their studies (see Figure ~\ref{fig:gpu-scaling}). The group needs to train deep Neural Networks for a large number of ensembles and nuisance parameters. Obtaining results for each iteration on a grid site with some handfuls of GPUs would take weeks or months, making it an unfeasible timeline for a PhD student. Purchasing hundreds of GPUs also does not make sense, since they will remain idle most part of the year.
\end{itemize}  

\subsubsection{CMS}
% Kenyi will talk to FNAL people. 
% Also want to include what was done on Amazon and Google a few years ago
% Approach is to not use it so much for large scale activities - primarily due to cost.

In an effort to extend the capacity of extensible resources to commercial cloud computing resources that the rental market has to offer, the CMS experiment started exploring this scenario back in 2016.

During this time, the Fermilab Tier-1 facility was expanded via HEPCloud~\cite{bib:cms-aws}, in order to dynamically provision commercial cloud resources coming from the Amazon Web Services and Google Cloud Engine~\cite{bib:cms-google}. 

\begin{figure}

\begin{subfigure}{.9\textwidth}
\centering
\includegraphics[width=\textwidth]{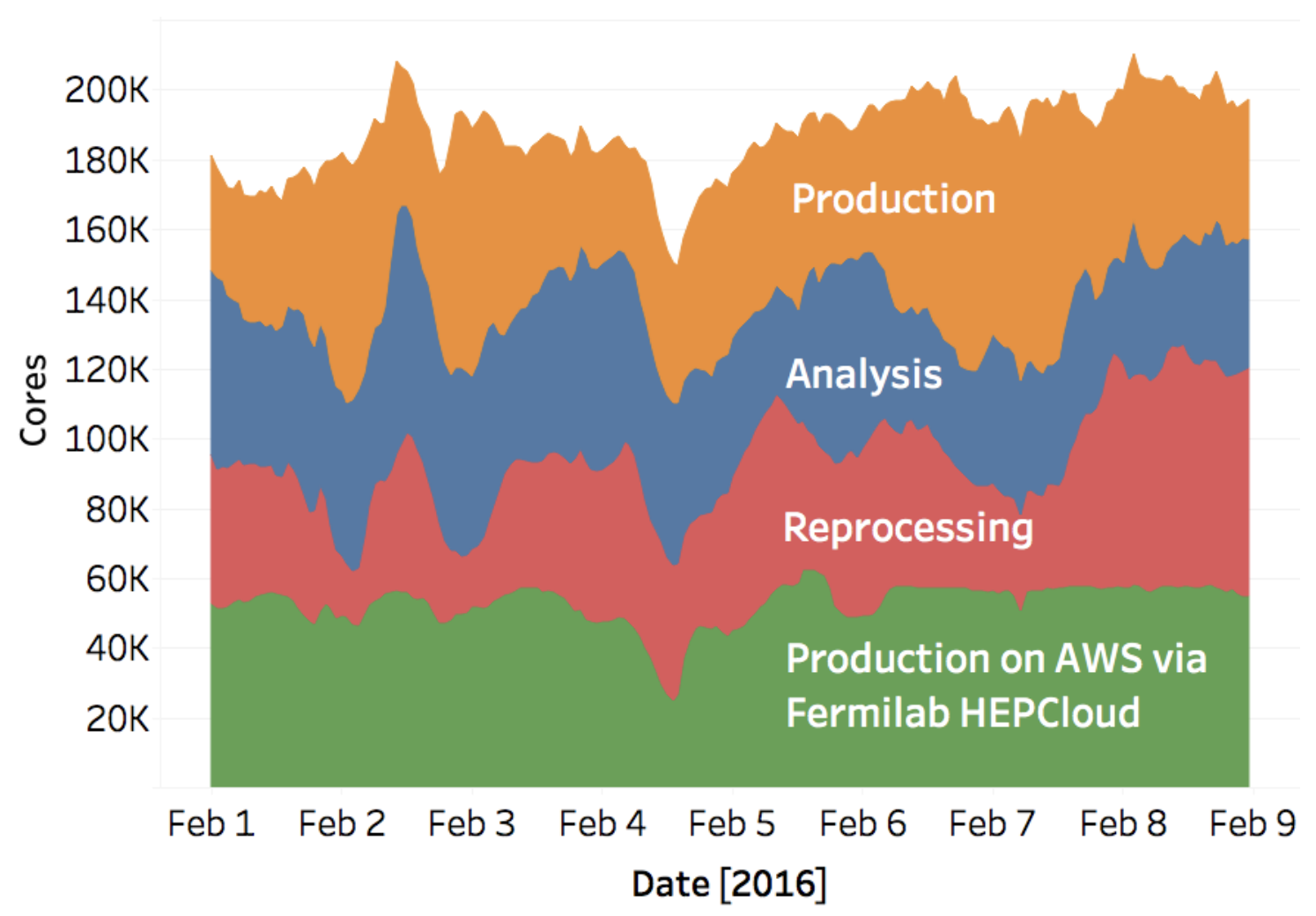}
\caption{A comparison of the scale of processing on AWS to other global CMS activity.}
\label{fig:cms-aws1}
\end{subfigure}

\begin{subfigure}{.9\textwidth}
\centering
\includegraphics[width=\textwidth]{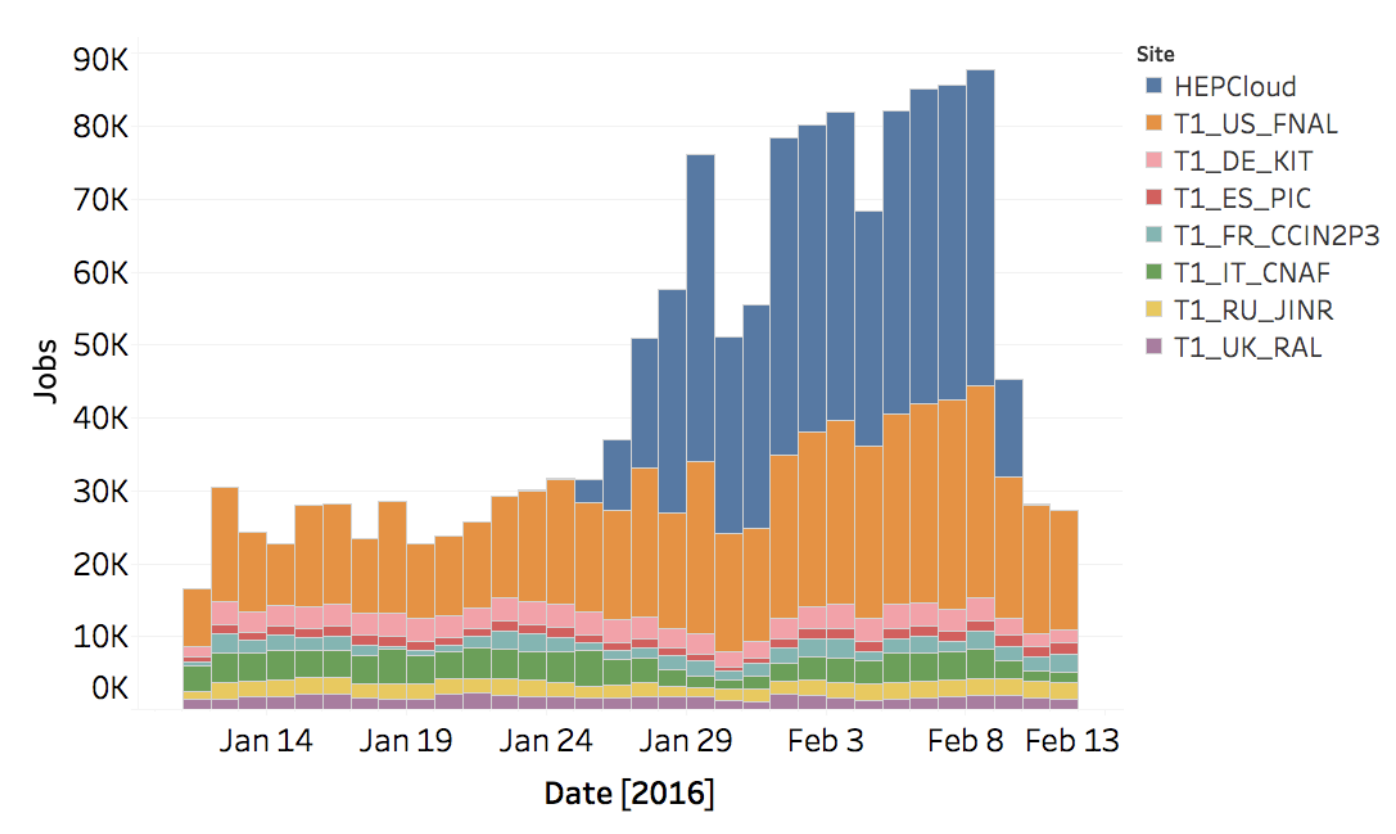}
\caption{A comparison of the scale of processing on AWS to other CMS Tier-1 activity.}
\label{fig:cms-aws2}
\end{subfigure}

\caption{Examples of cloud elasticity for resources not largely available at Grid sites. The cost of the resource usage depicted in each subplot is at the order of low, single-digit thousands of USD, but produced results useful at physics group or experiment level.}
\label{fig:cms-aws}
\end{figure}

\begin{figure}
\centering
\includegraphics[width=\textwidth]{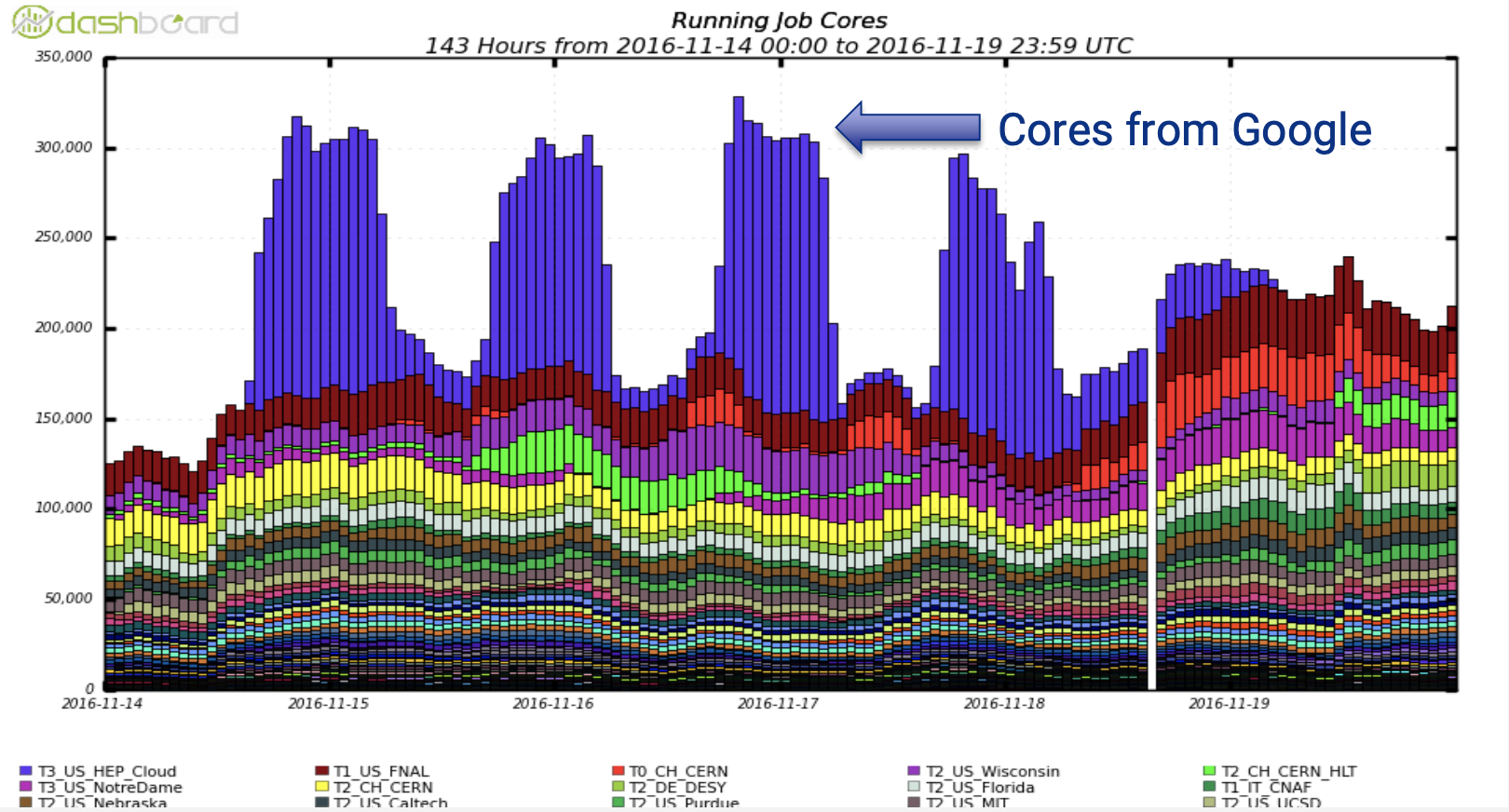}
\caption{A comparison of the scale of processing on Google to
other global CMS activity.}
\label{fig:cms-google} % Give a unique label
\end{figure}

Figures \ref{fig:cms-aws} and \ref{fig:cms-google} show a comparison of processing in AWS and GCE resources with respect to other Tier resources in the CMS Global Pool, consuming about 15 million and 6.35 million wallclock hours during these tests, respectively.

Production simulation workflows that chained together the GEN-SIM-DIGI-RECO stages were created in order to optimize the use of resources by encouraging longer running workflows with small outputs, to minimize data egress out of the cloud fees. The average CPU efficiency over all final job iterations found was 87\%, which is considered very good since these workflows ran every step of GEN-SIM-DIGI-RECO sequentially and not all these steps are
CPU-bound. The average efficiency for GEN-SIM, DIGI, and RECO on the grid are 57\%, 68\%, and 82\% respectively, for reference~\cite{bib:cms-aws}.

Lancium has also been integrated into production and validated with a few small workflows, even though the provisioning integration is still incomplete and a vacuum provisioning method is used at present (containers are started manually on demand).

\begin{figure}
\centering
\includegraphics[width=\textwidth]{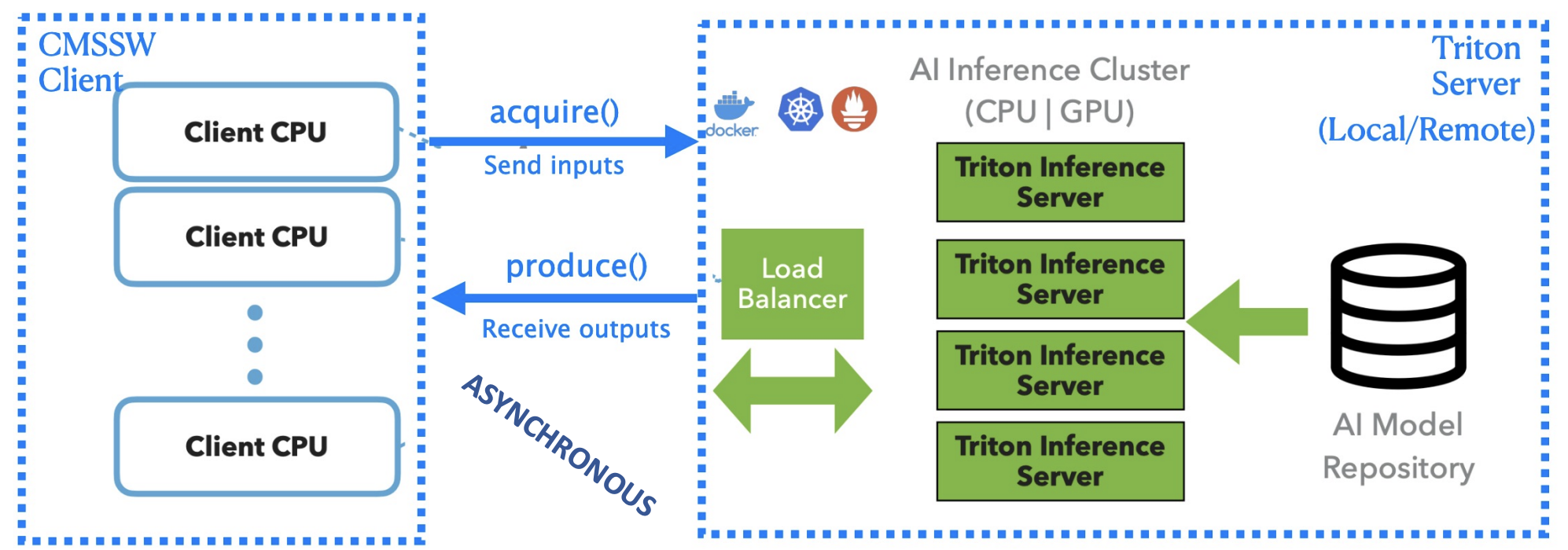}
\caption{CMSSW SONIC for Triton Inference Server diagram}
\label{fig:cms-sonic} % Give a unique label
\end{figure}

%Mention SONIC briefly, since we added this to the workshop slides
Cloud resources are also getting used for specialized analysis workflows. Figure \ref{fig:cms-sonic} shows a diagram of the usage of Services for Optimized Network Inference on Coprocessors (SONIC) to access  GPUs and CPUs in the cloud (e.g.: Google Cloud) for the CMS data analysis pipeline involving machine learning inference. The acquire() and produce() functions are used to send and receive information to and from the inference server connected to these cloud resources~\cite{bib:cms-sonic}.

\subsection{Cloud Integration}
\subsubsection{ATLAS} \label{cloud-atlas}
% Compute integration

There are many ways to integrate cloud resources with the ATLAS computing infrastructure. Sites will typically try to expand their on-premises resources, while the experiment or other entities could be interested in running an independent site on the cloud. This section will focus on the latest integration model that ATLAS followed to create standalone cloud sites for the Amazon and Google projects.

\paragraph{Data management integration} 
Rucio and the WLCG data transfer middleware (FTS ~\cite{bib:fts}) integrate cloud storage through signed URLs and the \textit{http} protocol, which are supported by many Object Stores in the cloud. A service account key is downloaded from the cloud provider and installed in the Rucio and FTS servers. Rucio and FTS will use the key to generate short-lived, signed URLs with permission to read or write a specific file. For direct file download or upload, Rucio provides signed URLs to the PanDA pilot or end users to interact with the storage with fine grained permission. For third party transfers, FTS provides the signed URL to the other storage element. The other storage element uses the signed URL to push or pull the file from the cloud storage.

As described in the section \ref{CAs}, cloud providers do not belong to the IGTF ~\cite{bib:igtf} and Grid storage elements do not trust certificates signed by the cloud Certificate Authorities, causing failed transfers. As a workaround a proxy element with an IGTF trusted certificate needs to be placed in front of the cloud storage. This proxy element needs to offer a very high bandwidth in order to not become the bottleneck for data transfers. The current solution is to load a CERN host certificate to a proxy element in the cloud. The measured throughput in our tests at Google was around 5GB/s, which falls a bit short compared to a large US site and needs to be watched during the operation of the ATLAS Google Site.

\paragraph{Workload management integration} 
PanDA has integrated cloud compute resources by direct job submission from their workload management system to Kubernetes ~\cite{bib:kubernetes} ~\cite{bib:harvester-kubernetes} clusters. Kubernetes is an open source system for managing containers across a cluster of nodes. While Kubernetes is most commonly used to manage services, it also provides batch controllers for job execution. ATLAS workload management system PanDA can communicate with Kubernetes clusters through the resource-facing component Harvester. This integration model is general to any Kubernetes cluster and avoids any vendor lock in. We have used the integration model extensively on various research clouds (CERN, University of Chicago, University of Victoria) and on commercial clouds (Amazon EKS, Google GKE, Oracle OKE). While we did not have the opportunity to try the Microsoft Azure cloud, we expect it will work as well. 

The only thing that needs to be installed on the Kubernetes cluster is the CVMFS file system. Due to certain restrictions on commercial clouds, the most suitable installation method is the usage of a CVMFS plugin for Kubernetes. At the beginning we had difficulties to run the plugin reliably, but we have been able to tune and optimize it during the proof-of-concept phases.

One of the most interesting features of managed clusters on commercial clouds is the auto-scaling capability. The cluster can scale in and out within predefined limits, according to the amount of pending jobs. When there are no jobs to process, the cluster will scale in to a minimum size and will avoid costs. When there are many jobs to process, a single cluster can scale out to a very large size. Given our current experience at Google, we are fairly comfortable scaling out clusters up to O(10\textsuperscript{5}) cores for periods of a day. This elasticity can be exploited to quickly process urgent payloads in the cloud.

We try to use \textit{spot} instances ~\cite{bib:gcp-spot}~\cite{bib:aws-spot} wherever possible for cost optimization and to evaluate the induced failure rate.

\paragraph{Additional services} 
The last component we need for an independent cloud site is a Frontier Squid service. The Frontier Squid caches metadata (ATLAS conditions data) required by ATLAS jobs and the CVMFS files. Like this, repetitive information does not need to be pulled through the Wide Area Network for each call. The Frontier Squid is a reliable service and does not require a lot of attention. On Google we have installed it through a Managed Instance Group. You can define the template for a Virtual Machine (e.g. operating system, packages and a startup script) and the number of desired, load-balanced instances. Automated health checks will test the service is running and replace unresponsive machines. On smaller clusters at Amazon, we install Frontier Squid as part of the Kubernetes cluster.

% https://monit-grafana.cern.ch/d/000000696/job-accounting-historical-data?orgId=17&var-bin=1h&var-groupby=processingtype&var-country=All&var-federation=All&var-resources=All&var-tier=All&var-cloud=All&var-site=All&var-computingsite=GOOGLE_BULK&var-nucleus=All&var-cores=All&var-eventservice=All&var-groups=All&var-inputdatatypes=All&var-inputprojects=All&var-outputproject=All&var-gshare=All&var-resourcesreporting=All&var-processingtype=All&var-jobtype=All&var-prodsourcelabel=All&var-jobstatus=All&var-error_category=All&var-container_name=All&var-job_resource_type=All&var-es_division_factor=1&var-pledges=CPU&from=1648959400430&to=1651587300707

\subsubsection{CMS} \label{cloud-cms}

In a similar way HPCs were integrated into the GlideinWMS Factory Pool via HEPCloud, the experiment also chose to use this technology to utilize cloud resources to run production workflows using AWS and GCE resources during the Fermilab HEPCloud pilot project, back in 2016. The main difference is that rather than creating new factory entries in the pool (that would make a new resource look like a new grid site), the Fermilab Tier-1 facility was expanded instead, since the focus was to dynamically provision cloud resources temporarily, to sustain the execution of many CMS workflows at an extremely large scale. For normal production integration, same restrictions as HPC apply in terms of storage for workflows matching. During the production workflow runs, the DIGI stage of the workflow was adjusted to retrieve input data files to the local worker node storage and perform the read operations from there, in order to decrease I/O cost related fees.

It is important to note that, while these production tests were executed in 2016, the infrastructure developed to make this happen is maintained to present, with the occasional development efforts that have been required in order to keep it up to date with the GlideinWMS infractructure. For example, supporting IDTokens authentication for AWS and Google VMs on the GlideinWMS side.

Another resource that has been integrated in the experiment is Lancium. This resource offers a much more reduced service catalog than major clouds, basically only allowing their users to submit batch jobs to the cloud and has no other services or storage options. In exchange, it also has lower list prices. Lancium has been cooperating with HEP experiments to make the integration and usage easier. For example, they have installed the CVMFS file system and Squid on their servers. Another difference between Lancium and AWS or GCE is that the former can only run singularity containers rather than VMs, so nested containers are used; the CMS pilot runs inside a singularity container and then the pilot runs singularity payload containers. Data input is read via AAA and staged out to FNAL. Due to the fact Lancium's API is not compatible with AWS or GCE, the provisioning integration is still incomplete; pilot containers are started manually on demand when needed.

Analysis workflows involving machine learning inference are currently testing the use of cloud resources by integrating machine learning inference servers in the CMSSW framework~\cite{bib:cms-sonictriton}. In particular, the use of the Nvidia Triton Inference Server via SONIC allows the deployment of these workflows in a variety of computing contexts, including GPUs and CPUs from dedicated Tier resources, HPCs and cloud resources. Current scale tests with this method used a mix of 100 GPUs and 10,000 CPU cores from Google Cloud~\cite{bib:cms-sonic}. This shows the flexibility that cloud gives us, being able to get many GPU nodes on demand on short notice and for a limited time.

\subsection{Cloud Computing Cost}
% No direct comparisons - probably.
% We need to do the work collecting information now, but we should write this last. It is going to be political
% Direct vs Indirect costs
% Direct
%    - hardware
%    - people
% Indirect
%    - other parts of the facility impacted by HPC/Cloud

% We also need to do costs of R&D and longterm operations.

% We should do costs for today - TCO for Cloud and HPC.
% Let's also extrapolate various scenarios and their cost impacts. This will be partially informed by the workshop.

In this section we will give an overview of our experience and lessons learned during our cloud activities. Note that this is limited to few providers (notably Google and Amazon) and might not represent the situation in other clouds.

\subsubsection{Cost components}

\begin{figure}
\centering
\includegraphics[width=\textwidth]{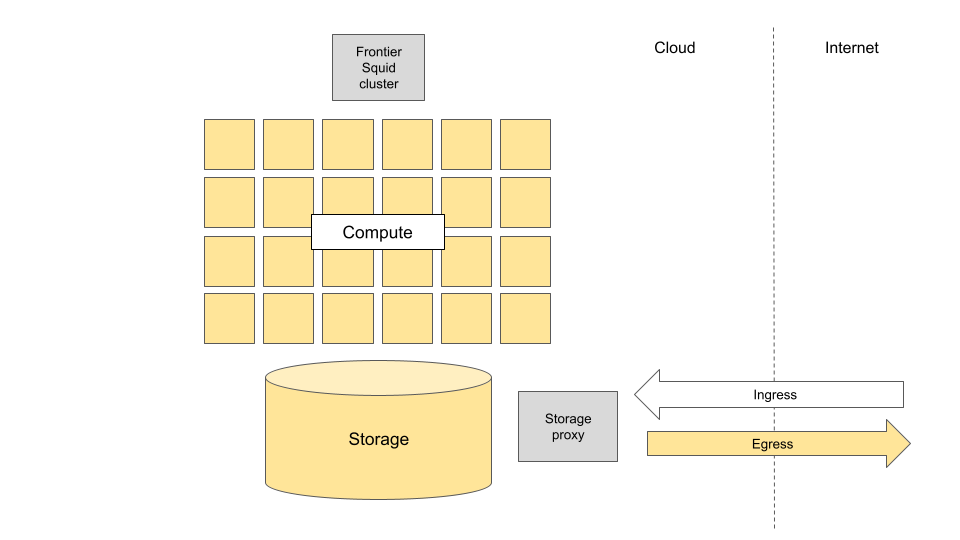}
\caption{Cost components}
\label{fig:cost-components} % Give a unique label
\end{figure}

A high level overview of the components needed for a minimalist, but fully functional ATLAS site can be seen in figure~\ref{fig:cost-components}.

Our main cloud cost components can be split into: 
\begin{itemize}
    \item Storage: the Rucio Storage Element, e.g. in the form of a cloud Object Store bucket.
    \item Compute: the PanDA queue with the worker nodes. If we implement the queue through Kubernetes, the main cost will come from the underlying Virtual Machines composing the cluster and a relatively small fee for the management of the Kubernetes cluster. If the cluster is sufficiently large, the Kubernetes management cost can be ignored. The cost of the Virtual Machines is the same whether it is part of a Kubernetes cluster or not.
    \item Network: data transfers to other sites. 
    \begin{itemize}
      \item Ingress: importing data into the cloud is free.
      \item Traffic within the same Region is free as well, so we need to build our cloud infrastructure consistently (i.e. put storage and compute in the same Region) and use internal IP addresses.
      \item Egress: we will be charged for the data sent to other Regions in the cloud and, most importantly, outside of the cloud. The traffic out of the cloud can be a substantial cost.
    \end{itemize}
\end{itemize}    
    
We will also need to set up a few auxiliary components:
\begin{itemize}
    \item Frontier Squid cluster: this is a minor cluster of small Virtual Machines.
    \item Storage proxy: this can be implemented as a Cloud Load Balancer.
\end{itemize}
The cost for these auxiliary components adds up to few hundreds of dollars per month and we are excluding them for the purpose of this document.

In Appendix ~\ref{cloud-componen-overview} we will try to guide through choice decisions of compute, storage and network components, their Qualities of Service and pricing models.

\subsubsection{Subscription Agreements for Public Sector}
For the current ATLAS Google project we have entered for the first time under a Subscription Agreement for the Public Sector, which seems to be unique to this cloud provider. This licensing option fixes the cost for any Google Cloud project. The Subscription Agreement removes most risks and uncertainty associated with properly sizing or unforeseen costs during the project. The terms and pricing are negotiated on a case by case basis and can provide substantial discounts. The subscription agreement is at this moment a way for both parties to get to know each other. At the end of the defined period, the usage will be reviewed and the terms renegotiated. 

In the ATLAS Google case, the terms of the contract were to use, \textbf{in average}, 10k vCPU and 7PB of storage over 15 months. The cost for this subscription is negotiated at the beginning of the project and, in our case, paid monthly. Any associated costs like egress are included in the agreement and we do not have to worry about them. It is the experiment's responsibility to make full usage of the agreed resources.

We have not yet renegotiated the contract and there is no guarantee that the terms will stay the same over the years. Experiments need to assess the risk for a potential price hike and have mitigation plans in place. This is true for any discount agreements.

\subsubsection{Operational effort}
Cloud computing generally has a higher cost for the infrastructure, but absorbs a large part of the operational effort and significantly speeds up development through the availability of advanced development frameworks with a few clicks. Through economies of scale and their large teams, Cloud providers can offer service qualities that are simply not attainable by a university or laboratory with limited staff. Most of the infrastructure in the cloud is autohealing and faulty elements are replaced transparently, so it doesn't need much attention.

For the Google ATLAS site, the ideal team profile has:
\begin{itemize}
    \item cloud computing experience
    \item DevOps mindset, i.e. the person is able to combine operations and development
    \item experience in the experiment's computing infrastructure, ideally data management and/or workload management
\end{itemize}
In practice, the operation of the ATLAS Google site is fully done by 0.1-0.2 FTE of a Rucio (data management) engineer and a PanDA (workload management) engineer.

For CMS, the operational effort to maintain the HEPCloud infrastructure that can integrate AWS and GCE in the  GlideinWMS pool requires about 0.1 FTE, with some occasional development efforts. For example, two to three months of work were required in 2022 to support IDTokens authentication for AWS and Google VMs on the GlideinWMS side.

The required person-power effort does not increase proportionally to the amount of resources, i.e. the same person can take care of a very large cluster. 

We need to also add some fraction of an FTE for the managerial effort to negotiate contracts with the Cloud provider and liaise with the other stakeholders (e.g. funding agencies and experiment).

\subsubsection{Rough budgeting of a cloud site} \label{CloudCostSimulation}

This section does not pretend to provide a price guideline and should not be used as a pricing source. Prices change over time and each project needs to decide the exact configuration of the resources.

Instead we want to sketch the cost of an ATLAS Google site according to different pricing options for compute classes. We do our exercise with Google, because we have the most experience on it and know this setup works. We do not seek to influence the choice of cloud providers and therefore avoid comparing prices with other providers. In particular, price comparisons across providers need to be done very carefully, making sure the products have equivalent disk and CPU performance.

Our site is configured to have:
\begin{itemize}
\item 10’000 virtual CPUs with 20 GB of scratch disk per core (current ATLAS recommendation)
\begin{itemize}
\item The virtual machines belong to the \textit{n2} machine series ~\cite{bib:n2series}, based on Intel's Cascade Lake or Ice Lake processors, and have 4GB of memory per virtual CPU.
\item The scratch disks are standard zonal persistent disks ~\cite{bib:standard-pds}, which corresponds to lower-throughput disks.
\end{itemize}
\item 7 PB of standard storage
\item 1.5 PB and 4 PB of Internet egress per month. These are the amounts of egress that were observed on US ATLAS sites of similar sizes at two different times of the year. It is very important to note that the amount of egress, which can not fully be controlled by the site, can become a dominating factor for the overall cost of a site.
\end{itemize}

\begin{figure}
\begin{subfigure}{.9\textwidth}
    \centering
    \includegraphics[width=\textwidth]{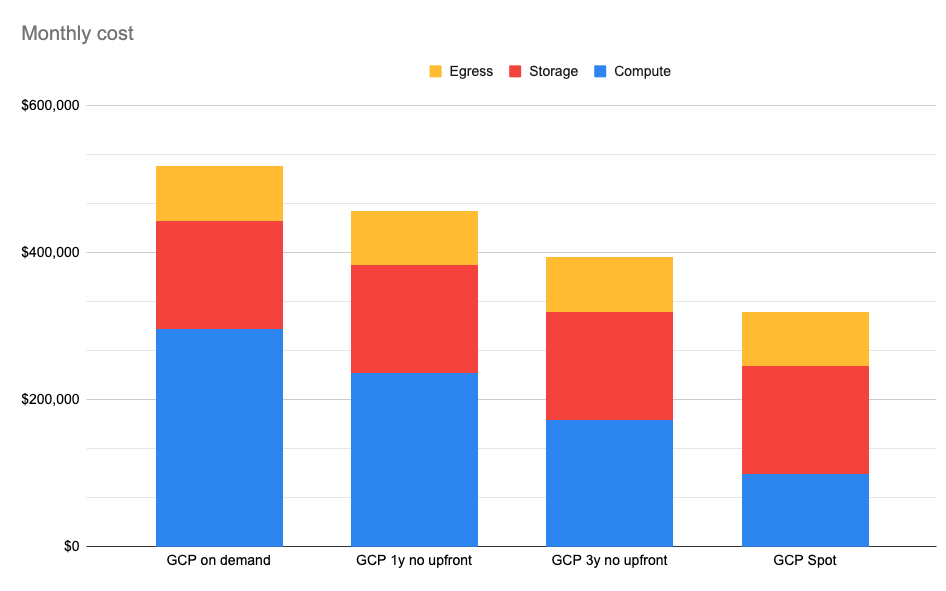}
    \caption{Monthly cost with 1.5PB egress}
\label{fig:monthly-cost-low-egress} % Give a unique label
\end{subfigure}%

\begin{subfigure}{.9\textwidth}
    \centering
    \includegraphics[width=\textwidth]{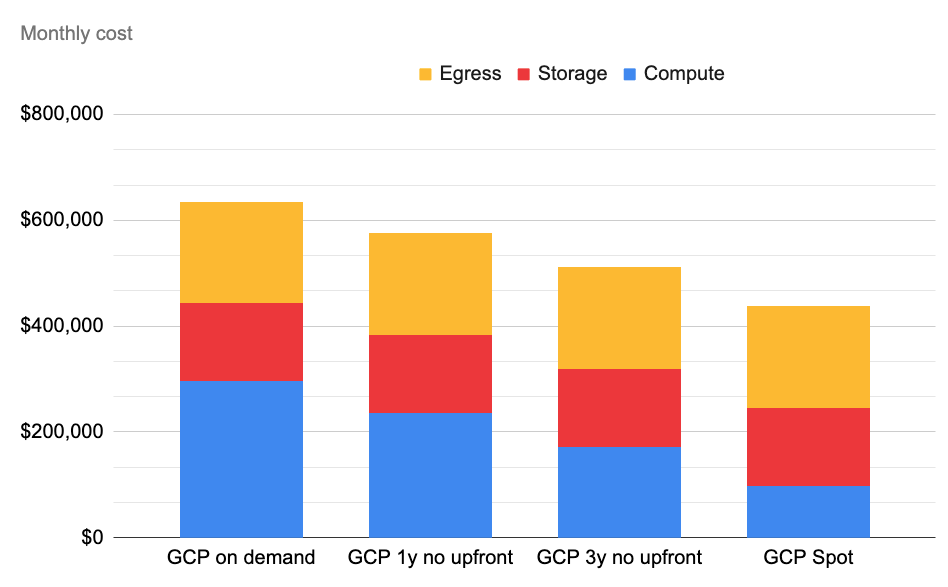}
    \caption{Monthly cost with 4PB egress}
\label{fig:monthly-cost-high-egress} % Give a unique label
\end{subfigure}
\caption{Monthly cost of the ATLAS Google site according to different pricing models. We only have access to 3 months of data transfer volumes in Rucio and these were the two scenarios we have observed at different times of the year. The two estimations with different egress volumes demonstrate how the egress can become an unpredictable, dominating factor for the overall cost of a site. 
The subscription is significantly lower, but not included in the diagram since it was negotiated for our particular case and might or might not apply to other projects or in the future. There are also possibilities to negotiate certain discounts on the egress.}
\label{fig:monthly-cost} % Give a unique label
\end{figure}

Figures  ~\ref{fig:monthly-cost-low-egress} and ~\ref{fig:monthly-cost-high-egress} show the list-price comparison for the site with \textit{on-demand}, \textit{reserved} (with 1 year and 3 year commitments) and \textit{spot} Virtual Machine instances. In our first months of usage, and at the scale of our site, we have been very satisfied with the \textit{spot} instances. The induced failed wall-clock by \textit{spot} is at the low, single-digit level, while the price is significantly lower (around 60\%) than \textit{on-demand} instances.

In practice, the customer should leave some overhead for incidental costs like network devices (e.g. NAT for private clusters), log storage (e.g. the \textit{stdout} in the cluster can be redirected to a log storage for convenient examination) and others. These incidental costs can be ignored in the subscription model.

It is frequently debated in the HEP community whether large cloud customers really pay list prices and how heavily prices can  be negotiated. Services and products are generally cheaper when bought at bulk rather than retail, and this is also the current experience in our negotiations with cloud providers. In fact, our current subscription offers the services at a fraction of the list price. Therefore we consider that both the list price and the discounted or subscription price are relevant. As long as substantial discounts are available, the cloud can be considered for the provisioning of complimentary resources for bulk processing.

\subsubsection{Cost Estimation during CMS Production Tests} \label{CMSCostEstimation}

CMS production tests with AWS and GCE spot instances in 2016 resulted in estimated total costs of 1.4 and 1.6 cents per core-hour, respectively, which was close to the 0.9 cents per core-hour for on-premises resources~\cite{bib:cms-aws} \cite{bib:cms-google}. While this is useful to show, it is important to note these numbers are outdated at present and can't be used as a guideline, as spot instance prices are changing.   

\subsubsection{List prices for ARM and GPU exercises} \label{ExoticCloudCost}
We would also like to show the list prices for the smaller ARM and GPU exercises, which came as by-products of the ATLAS projects on Amazon and Google. Given the already existing cloud experience, the setup of the resources took some days of work. From a person-power perspective, it would have been more difficult and time consuming to run these exercises as completely independent projects and with separate, possibly inexperienced teams.

\paragraph{Amazon ARM queue} 

We created an ARM queue to run the Physics Validation of the Athena simulation software. Two attempts were needed and the simulation tasks were executed within few days. The overall list price for the exercise was 4.1k USD. We do not know if this is less than the cost of operating an on-premise ARM cluster, but it is certainly much less than the cost of buying a new ARM cluster. The Physics Validation included only the Simulation software and not the other parts of Athena. The validation of the Athena simulation software for ARM is the first necessary step towards having ARM payloads. This could offer future possibilities to use certain HPCs or evaluate whether ARM CPUs are more cost effective than x86 for on-premise resources. \\

\begin{figure}
\begin{subfigure}{.9\textwidth}
  \centering
  \includegraphics[scale=0.4]{ATLASCloud/FRESNO-ARM_validation.png}
  \caption{Resources used.}
  \label{fig:google-gpu-scaling}
\end{subfigure}%

\begin{subfigure}{.9\textwidth}
  \centering
  \includegraphics[scale=0.3]{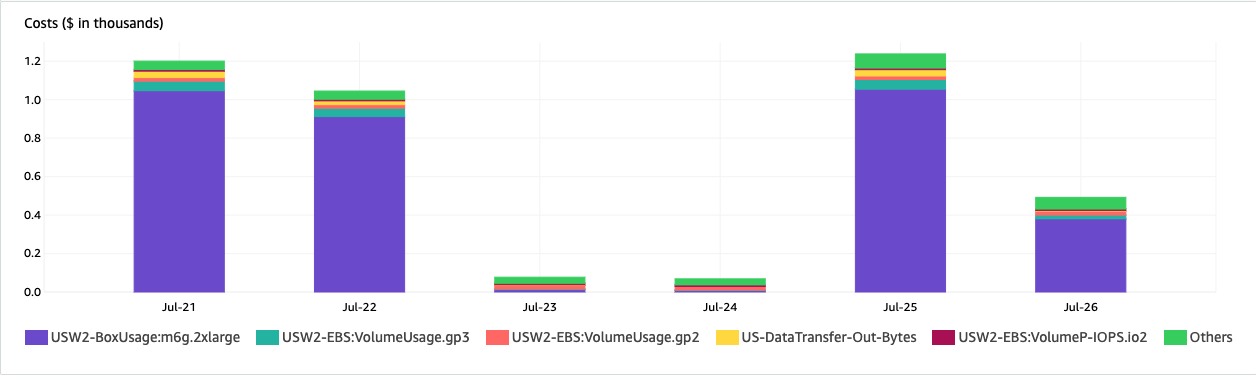}
  \caption{Cost seen in the billing console, which reflects list prices. We do not know if the CSU Fresno credits were purchased at a reduced price.}
  \label{fig:google-gpu-costing}
\end{subfigure}
\caption{Resources and cost for the \textit{arm64} Physics Validation.}
\label{fig:google-gpu-exercise} % Give a unique label
\end{figure}

\paragraph{Google GPU queue}

We have been operating a GPU (Nvidia T4) batch queue under the ATLAS Google project and made it available to a collaborating physics group. Figure ~\ref{fig:google-gpu-exercise} shows the last 3 months (August to October 2022) of resource usage and the daily cost as viewed on the billing console. The billing console reflects the list price, not our real cost. The resource usage is accounted in cores (4 CPUs per GPU), so we have been using between 15 GPUs in August and 200 GPUs in October. When the resources are not used, they are returned automatically to Google and avoid any cost. The cost for the 3 months according to list prices has been 3.4k USD. The access to such a GPU pool has enabled the work of a PhD student and their physics group. From an infrastructure point of view and considered as a completely independent project, it would require a fraction of a person to set up and operate the GPU queue. As part of the ATLAS Google site it is done with little additional effort. Purchasing 200 GPUs would require an front up investment of some 100k USD and would make little sense, since there are not enough users for continuous usage.

\begin{figure}
\begin{subfigure}{.9\textwidth}
  \centering
  \includegraphics[scale=0.4]{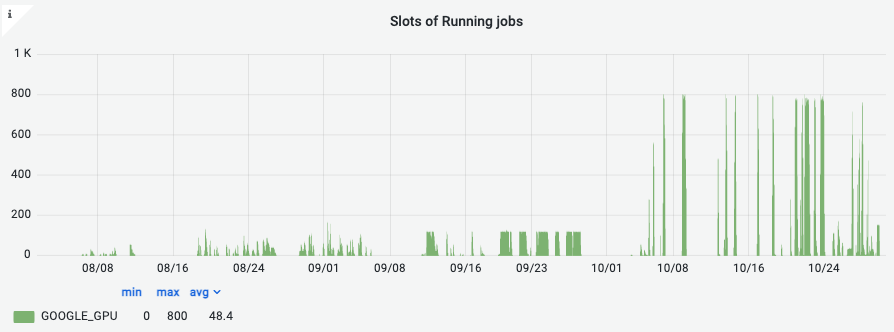}
  \caption{Resources used. Each node has 4 cores (shown in the y-axis) and 1 Nvidia T4 GPU.}
  \label{fig:google-gpu-scaling}
\end{subfigure}%

\begin{subfigure}{.9\textwidth}
  \centering
  \includegraphics[scale=0.35]{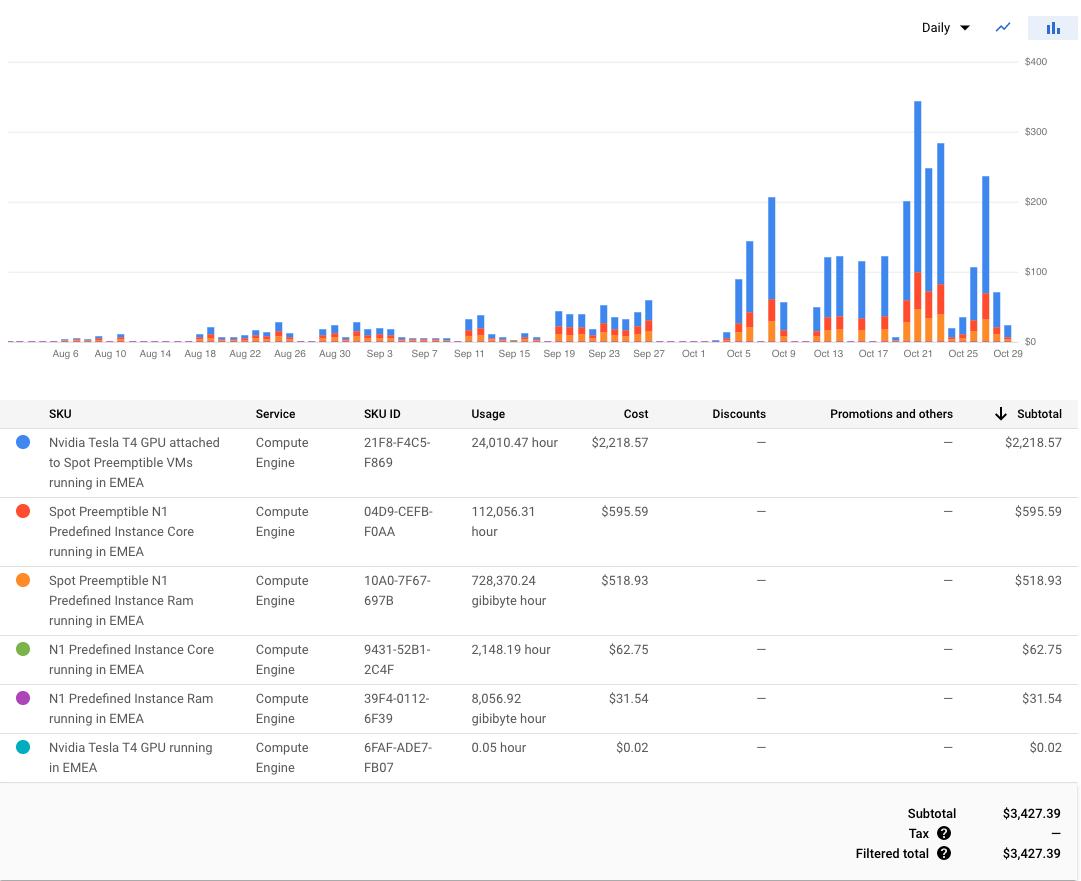}
  \caption{Cost seen in the billing console, which reflects list prices, not the real prices. GPUs only run when they are in use, otherwise only a small CPU node needs to run to keep the Kubernetes cluster alive.}
  \label{fig:google-gpu-costing}
\end{subfigure}
\caption{Resources and cost for the "Off-shell Higgs Measurement Using a Per-Event Likelihood Method" by one of the physics groups at the University of Massachusetts Amherst.}
\label{fig:google-gpu-exercise} % Give a unique label
\end{figure}

\subsection{Cloud Computing Concerns} \label{Concerns}

We would like to briefly go through reappearing concerns.

\begin{itemize}
\item Data custodiality: given the current, short-term contracts with cloud providers, clouds should only be used for the storage of transient data that is also available on other Grid sites or simulated data that can be generated again. If the experiment wants to use cloud storage for custodial copies of data, it would be recommendable to shield the conditions contractually and ensure there is a strategy to copy the data to another storage on time. It could even become a larger problem in the event that the cloud provider runs out of business.
\item Loss of expertise at the sites and operational effort: there is the worry of losing technical person-power at the sites that also participate in daily computing operations of the experiment. The counter argument is that the usage of cloud computing would allow physicists to focus on science, instead of branching out to computing activities.
\item Vendor lock-in: worry about getting locked into a provider-specific platform and making the exit complicated. It is important to choose technologies generic to most cloud providers and we have done so in most integrations.
\item Loss of ownership of resources: cloud computing resources are only available while they are rented. At the end of the contract there is no hardware available.
\end{itemize}

\subsection{Further Cloud Computing R\&D}
\subsubsection{Full chain or other ways to reduce egress costs} \label{full-chain}
To date, ATLAS is starting to explore ways to reduce egress costs. The current idea is similar to the CMS \textit{StepChain} concept explained in section~\ref{CMS-HPC-Workflows} and called \textit{Full chain} in ATLAS. The objective of \textit{Full chain} is to run all the steps in the Monte Carlo production chain consecutively in the cloud or any other particular resource. The intermediate data products are kept in the cloud and only the final product is exported. A chain could be executed very quickly and little egress would be generated.

\subsubsection{GPU, TPU, FPGA and other resources}
The cloud is an ideal place to carry out various R\&Ds and gather experience with expensive resources. We tried to show a few examples in sections ~\ref{ExoticCloudCost}, where the availability of different types of hardware allowed us to carry out powerful activities at medium scale without incurring in a large upfront investment. The experiments have little experience with many of these resources both from the user and from the infrastructure point of view. The integration of these resources into the experiment frameworks and how to make them available \textit{democratically} needs further research.

\subsubsection{Upper service levels: Software as a Service, AI as a Service}
Clouds offer hosted services that can be evaluated and compared with the experiments' hosted services. Examples are hosted notebook services, distributed training platforms and even AI as a Service. The danger with some of these higher levels is to completely loose the know-how and completely depending on external sources.

\subsection{Cloud Computing Efforts Conducted Internationally and by Other Virtual Organizations}
We would like to provide some pointers to other efforts carried out by participating institutes and other Virtual Organizations.

\subsubsection{Expansion of the ATLAS T2 at the University of Tokyo}
% Tokyo University Google project: https://www.epj-conferences.org/articles/epjconf/pdf/2020/21/epjconf_chep2020_07034.pdf
The University of Tokyo has been looking into alternatives to expand their ATLAS T2 site through external cloud resources (Google) in early 2019 ~\cite{bib:tokyo}. 
In order to use Google compute resources, they used a similar architecture to the Fermilab HEPCloud used by CMS. They developed a tool to manage Virtual Machines acting as worker nodes that would join their HTCondor cluster at the University of Tokyo. The worker nodes would use the University storage for input and output.  They ran a cluster with up to 1k virtual CPUs for a few weeks and executed a variety of production jobs. They ran a brief cost study that has some differences, but is compatible with the numbers provided in our cost estimations using list prices (see Section \ref{CloudCostSimulation}) .

\subsubsection{European initiatives with participation from CERN: HelixNebula Science Cloud and OCRE}
% CERN and Cloud Bank: https://indico.cern.ch/event/948465/contributions/4324023/editing/paper/515/6032/2nd_Revised_version__vCHEP_2021_CloudBank_for_Europe.pdf

The European Union has been promoting public-private partnerships between research entities and commercial clouds in order to bolster the European IT landscape. The two better known projects with CERN's participation are HelixNebula Science Cloud and OCRE ~\cite{bib:ocre}. They intend to bring together research organisations, public infrastrucutre providers and commercial clouds. OCRE was conceived to provide easy, regulation-compliant procurement and single entry points to a variety of cloud providers operating in Europe.

CERN, in their technical participation in these projects, has extended their batch system into the cloud or provided access to accelerators using novel frameworks. Independently from our cloud activities, these projects also based their deployments on Kubernetes.

\subsubsection{Vera Rubin Observatory}
The Rubin Observatory developed their computing infrastructure on top of the Interim Data Platform (IDF) at Google Cloud,  while they worked on building out their own distributed computing infrastructure. The resources on the cloud were available with short delays, while developing their on-premise infrastructure is coming years later.

Their decision is to develop a hybrid model. Most storage and bulk compute will be hosted on-premise. Their interactive, user-facing Rubin Science Platform will be hosted on the cloud. For this purpose, they have recently signed a 3-year agreement with Google cloud. 

The Vera Rubin observatory shares similar worries as other experiments (avoiding egress costs), but did not seem to be worried about vendor lock-in, as they use generic frameworks, or about price hikes, as they rely on the competition between providers.

\subsection{Desirable Cloud Computing facility features and policies}
Besides a few improvements regarding integration with the Grid infrastructure (see Sections ~\ref{CAs} and ~\ref{LHCONE-integration}), cloud providers provide very solid technologies and infrastructure for the HEP community. The desirable policies revolve around cost and ensuring contractually the cost remains stable for a period that allows the experiments to manage risks and prepare contingency plans.

\begin{itemize}
\item The biggest concern that everyone expresses is egress cost. It is not known if the egress cost is an artifact to avoid customers exiting the cloud and exporting their data, or they are needed to cover the construction and maintenance of the cloud's network. 
\item The duration of our contracts so far has been short (between 1 to currently 15 months) and the providers have offered attractive deals, either by matching credits or through a very favorable Subscription Agreement. There is no experience renegotiating these contracts over a longer period of time. The cloud is therefore relegated to a secondary role, in particular with no experiment trusting them the custodiality of data. In order to progress with the relationship, the cost evolution in the longer term would have to be shielded contractually.
\end{itemize}

\section{Further integration work for external resources}

\subsection{Pledging} \label{Pledging}
Collaboration with WLCG is typically formalised through a Memorandum of Understanding (MoU). The MoU sets up the framework of agreement regarding resources and services that a site will provide to an experiment. Sites formally pledge a certain number of normalized core and a commensurate amount of storage for a certain period. The formal association allows the experiments to plan their campaigns and computing model in exchange for giving credit to the sites.

With few exceptions, the integration of HPC and Cloud resources does not follow the formal approach and these resources are labelled \textit{opportunistic} or \textit{beyond-pledge}. While being very welcome by the experiment and providing additional capacity, \textit{beyond-pledge} resources are difficult for planning purposes. They are not necessarily available when the experiment needs and, in particular, HPC scheduling policies are not compatible with the MoU requirements, since the queuing time in the HPC is not controlled by the submitter. In many cases, beyond pledge resources do not have the optimal architecture for HEP computing and not all workflows can be executed. This raises the debate how to compare the credits received by a fully compatible site that can run any type of job at any given point of time, with an opportunistic resource that only runs some less demanding jobs at a not fully-controlled point in time.

The discussion becomes more complicated when we extend it to GPU resources, for which the experiments currently have no pledging framework.

The pillars for pledging are benchmarking (measuring the power of the CPUs) and accounting (accounting and reporting the CPU usage), which will be discussed in the next sections.

\subsection{Benchmarking} \label{Benchmarking}
The CPU pledges use normalized core power according to the HEPSpec06 benchmark and is not measured in the number of generic cores. 

This means that for a particular compute cluster with more than one CPU model, all CPU models need to be benchmarked and averaged out. When using cloud resources this can become more tedious, since providers can mix multiple CPU models in a Virtual Machine family and possibly mask the model information. This will lead to some error margin on Cloud accounting.

The next benchmark generation will be HEPScore. Their effort to build containerized benchmark tools should simplify the measurement of HPC and Cloud hardware.

While some effort exists to benchmark GPU resources, the variation in architectures will make it increasingly complicated to compare them among each other. This is equally applicable to both Clouds and HPCs.

\subsection{Resource accounting} \label{Accounting}
The consumed CPU time is measured and reported periodically to the WLCG accounting framework by the Compute Elements, developed as part of WLCG middle-ware.

Some of the integration models described in previous sections, in particular for ATLAS, bypass the Compute Element and connect the batch system directly to the experiment's Workload Management System. In these cases, an alternative method to report the accounting numbers to WLCG needs to be implemented. Individual solutions exist and are in use by University of Victoria ~\cite{bib:kapel} or BOINC. These solutions should be extended to a generic service, where resource specific or Workload Management System based plugins need to be added.

Although some low-level tools exist for the accounting of GPU resources, there is currently no experience with GPU accounting in the experiments and they are not included in the experiment's frameworks. 

\subsection{Certificate Authorities} \label{CAs}
% Cloud inclusion in IGTF or accept other CA certs 
% Otherwise limited bandwidth to cloud storage to ~5Gb/s
Certificates used in the WLCG are provided by the Interoperable Global Trust Federation (IGTF) framework. Intercommunication between sites relies on certificates signed by an IGTF Certificate Authority (CA). Cloud providers' CAs are not part of IGTF and consequently Grid services do not trust these certificates. This is an issue for third party transfers scheduled by Rucio/FTS between a Grid storage element and a Cloud storage element. Workaround solutions need to be put in place, which do not provide the same performance and limit the transfer bandwidth. WLCG has put in place a task force to find a solution, but there is no immediate timeline foreseen to provide a solution.

% https://indico.cern.ch/event/876793/contributions/4515722/attachments/2305473/3922744/GDB-FutureOfHostCertificates-2021-09-07-2.pdf

\subsection{Network integration} \label{Network}
Large LHC data transfers should ideally be routed through private networks and avoid the public internet to guarantee acceptable latency, bandwidth and overall Quality of Service. This happens through the various national research networks, which are interconnected worldwide into the LHCONE and LHCOPN networks.

\subsubsection{Site-to-site peering} \label{Peering}

In the US, the Energy Sciences Network (ESNet) serves all DOE National Laboratories and already provides connectivity to DOE HPCs and multiple commercial clouds. 

ESNet6 has built the physical network into each DOE national lab, many of which host large computing facilities important to ATLAS and CMS work. This includes the ATLAS and CMS Tier 1 facilities at Brookhaven National Laboratory and Fermilab respectively, the Leadership Computing facilities at Argonne and Oak Ridge, and the NERSC facility at Lawrence Berkeley National Laboratory which hosts one of the HPCs most effectively integrated to date. WAN connectivity for DOE HPCs is generally not a limiting factor and are well connected to other DOE labs participating in the ATLAS or CMS Grid infrastructure.

% I'm hesitating to write the NSF HPC summary: "Some very well connected, others not so much, depending on where they are"

As well, ESNet has already established multiple 100 Gbps interconnects to several commercial clouds in the US (Amazon, Google, Microsoft, Oracle). Depending on the particular case, the interconnect can be specific to a Region or nationwide. A similar interconnect exists between Google Zurich and CERN. For the supported clouds, data traffic with a DOE national lab participating in the ATLAS or CMS Grid infrastructure will automatically run over ESNet. 

\subsubsection{LHCONE integration} \label{LHCONE-integration}
Participation in the LHCONE network is based on IP address ranges that are added to the \textit{allowlist} and requires exclusively LHC data to travel through the network. These requirements make it a difficult task to integrate a cloud provider in LHCONE:
\begin{itemize}
\item How to limit the IP ranges for a cloud provider or one of its Regions? Even if known, the ranges can be too broad to be accepted. 
\item How to ensure that only LHC data is being transferred and not data from a different cloud service (videos, music)?
\end{itemize}

Concluding, reliable WAN connections exist between HPC/cloud and DOE labs and it is possible to exploit these for data transfers or associating them together. However overall integration between cloud providers and the WLCG does not seem like a short or medium term possibility. We also do not know if these interconnects have any impact on cost.

\subsection{Functions as a service} \label{faas}
% Also relevant for HPC, see: https://parsl-project.org/parslfest2021-files/feickert-pyhf.pdf

Functions-as-a-Service (FaaS) is a model for running scalable, event-driven microservices first pioneered in the Cloud space. The idea is that computing resources can be abstracted in such a way that developers can skip infrastructure provisioning steps (i.e., IaaS) by sending self-contained functions along with data to an orchestration service which dynamically provisions instances of an appropriate programming language runtime, processes the data with a given function, and sends back the results. In this way, developers can dynamically scale function evaluators based on demand and only pay for what they use. This idea has been extended into the HPC space with tools like FuncX~\cite{bib:funcx}, which provides a Python-based library for bundling and dispatching functions to a broad variety of HPC, HTC and Cloud resources. 

Members of both ATLAS and CMS have done some work with evaluating approaches like FuncX for use in end-stage analysis facility pipelines, and similar approaches to abstracting HPC batch interfaces are taken by the ATLAS Event Service and Raythena. As this model originated with cloud computing, it is a very good fit for that space as well. While FaaS would add new capabilities, the exact operations deployment models and total resource footprint are not clear at this time.

\subsection{Machine Learning / Artificial Intelligence}

We already mentioned the adoption of GPU/Accelerator resources at HPC and how they represent a paradigm change for our software that we can't simply fix with a rebuild. We need to redesign algorithms instead. Apart from these ongoing efforts, there are also efforts to use these GPU/Accelerator resources for ML/AI and then integrate these into how we are doing computing (i.e. replace algorithms with inference calls). This would then also be applicable to GPU/Accelerator resources that are available on Clouds.

We do this for two reasons. ML/AI models could lead to better Physics performance of our code than more traditional computational algorithms and/or they could lead to overall smaller resource usage (if the cost of training is smaller than the savings we get from replacing computational algorithms with inference calls). To first order we of course want both, since then the ML/AI approach is clearly an improvement. It's hard to see the experiments accepting a worse Physics performance from a ML/AI approach, but a better Physics performance at the cost of a larger resource usage is something that cannot be ruled out at this point.

ML/AI and how to use it in HEP Computing is a very active area of R\&D. There are promising first results, but the exact operational deployment models are still being worked out. It's also unclear how much of a total resource usage footprint this would represent, i.e. could we really use (and keep using over multiple years) large LCF allocations just for ML/AI? This will depend on how expensive training such ML/AI models is going to be (early studies show that hyperparameter optimizations can be very expensive) and how often we will have to train and retrain (i.e. how stable such training results will be assuming detector and data taking condition changes over time).

\section{Comments and recommendations} \label{comments-recommendations}
% Big topic: GPUs. How are we going to use them. 2 ways: Algorithmic and ML/AI
% In the short/mid term we can ignore HPC pledging. In the future, if HPCs become a large fraction of our total resources - then a big chunk of our resources are uncredited and unplanned from the pledging concept. Recommendation? Needs to be worked out with the experiments and WLCG. May be out of scope for this report. 

% We should say something about effort cost. R&D on integration and operation will be needed.

\subsection{Common comments and recommendations}

\begin{itemize}
    \item [Comment 1] The experiments feel comfortable in the HTC (High Throughput Computing) \textit{x86} world. Architecture evolution indicates a growing heterogeneous landscape with proliferation of different CPU architectures and widespread adoption of GPUs, in particular in HPCs. Computing models and software by more recent sciences were potentially designed with GPUs in mind, and are able to run on HPCs exploiting CPU and GPU resources. 

    \item [Comment 2] There are initiatives within the LHC experiments and the larger HEP community (through HEP-CCE for instance~\cite{bib:hep-cce}) to port our software and the underlying dependencies to heterogeneous architectures, but it is a very complicated endeavour to adapt an enormous software stack that was written over many years. There are also activities that invest in developing ML and AI applications. To date, these are not enough to fill a sizable, GPU-dominated HPC allocation.

    \item [Recommendation 1] We recommend US ATLAS and US CMS to develop WLCG accounting and pledging strategies for their opportunistic resources, particularly for the integration models that bypass a Compute Element, connecting the batch system directly to the experiment's Workload Management System.
\end{itemize}

\subsection{HPC}
\begin{itemize}
    \item [Comment 3] Today, sending workloads to HPCs is considerably easier than it has been in the past for three reasons: 
        \begin{itemize}
            \item The collaborations have invested significant amounts of effort in developing tools and pipelines for managing jobs and data at HPCs which are running in production for several years now.
            \item Containerization and CVMFS are either available or easily made available thanks to operating system advancements.
            \item Machine architectures are somewhat friendlier to our workloads, given the broad availability of \textit{x86} CPU-only partitions.
        \end{itemize}

    \item [Comment 4] It is important to note that even though machine architectures are somewhat friendlier to our workloads, this is not a point of long-term stability. The ever-increasing need for more FLOPS with constrained space, power and cooling implies that GPUs or other accelerators will be increasingly important in the future, meaning that CPU-only partitions may eventually be a quite small fraction of total capability at a given facility. It is also not necessarily the case that the CPU partitions will be provided via solely Intel/AMD for future machines, as HPC centers consider competitive bids from many vendors including those providing POWER or ARM architecture. 
    
    \item [Recommendation 2] For the reasons pointed out in Comment 4, US ATLAS and US CMS should invest in software flexibility as outlined previously.

    \item [Comment 5] Allocations in US HPC centers require competitive proposals in order to receive allocation time. There is a non-trivial amount of effort spent in writing proposals with strong physics use cases, reporting back to the facility with results produced by time spent on resources at the HPC and so on. 

    \item [Comment 6] The US CMS and US ATLAS experiments are still developing or validating GPU production code that could allow them to exploit LCF resources.

% Recommend that all US HPC facilities implement a common API for interacting with storage and compute

    \item [Recommendation 3] Work with US HPC facilities to standardize on a technology for providing a facility API that permits both interactive and automated access to manage job workflows and data, in order to integrate better with ATLAS, CMS and other large-scale multi-institutional scientific collaborations. If such an API were standardized, it would require some significant R\&D up-front to integrate with our respective workflow management systems but would ultimately reduce the costs of initial integration and ongoing operations in the long run.

    \item [Recommendation 4] Work with US HPC providers to develop a plan for providing large storage allocations that can federate with existing ATLAS/CMS data management software, such that HPC storage can be used in a manner similar to how WLCG site storage is used today (i.e., as Rucio Storage Endpoints).

    \item [Recommendation 5] Work with US HPC providers to develop a plan for providing multi-year allocations for storage and compute such that HPC resources can be incorporated into ATLAS and CMS planning horizons.

\end{itemize}
\subsection{Cloud}
\begin{itemize}
% Adjusting our planning to consider and take advantage of flexible scheduling for peak. What does it allow us to do that we can't currently do?
    \item [Comment 7] Cloud projects have gathered some inertia over the last few years and help to foster interactions between industry and science. They contribute to the knowledge transfer and help the experiments extend their expertise outside of the usual environment. 
    \item [Comment 8] Useful integrations and cloud setups have been demonstrated on major cloud providers at small and large scale. Integration ideas for novel cloud providers exist. 
    \item [Comment 9] Cost remains a concern, however current user subscriptions and discounts help bridge the gap and make medium scale computing on the cloud possible. There is some uncertainty on the evolution of prices and dependency on cloud providers. For this reason, cloud resources should be considered complimentary to grid resources, rather than an alternative.
    \item [Comment 10] We have tried to provide cost examples that can be used by sites, the experiments and funding agencies to help the decision whether to rent cloud resources and to find the balance between owned and rented. 
    \item [Comment 11] We consider the cloud an ideal substrate for carrying out various R\&Ds that require elastic usage of resources, accelerators or specialized hardware. Generic self-service access to the cloud, where e.g. the users have to configure their resources independently, is more suitable for people involved in computing activities. Physics-analysis users usually have less computing experience and tend to abandon if there is no tailored service. The multiple cloud projects have tried to engage users and find working combinations between users and computing experts.
    \item [Comment 12] Cloud projects need continuity, if they are abandoned for a period, it takes effort to restart and build up the experience again. Smaller initiatives (e.g. the tests on ARM or GPU), which have larger overheads to run independently, are often absorbed and done as part of the larger initiatives. Experience with these resources is strategic to the experiment.
    \item [Recommendation 6] US ATLAS and US CMS should reassess the total cost of operation of cloud platforms after completing the Google Cloud Platform and Lancium projects.
    
\end{itemize}

\section{Common R\&Ds} \label{common-rnds}
\begin{itemize}
    \item [HPC] From sections \ref{xrootd-service}, \ref{service-platforms}, and \ref{faas}: R\&D on adopting potential common remote execution APIs (see Recommendation 2)
    \item [HPC] From sections \ref{hpc-rucio-globus} and \ref{hpc-storage}: R\&D to make effective use of larger storage allocations (see Recommendation 3) 
    \item [HPC] From section \ref{hpc-integration}: R\&D to increase job scheduling efficiency
    \item [HPC] From section \ref{mfa}: R\&D on acceptable authentication for WFMS pilot submission etc
    \item [HPC] From section \ref{xrootd-service}: R\&D on HPC data transfer mechanisms
    \item [Cloud] From section \ref{special-cloud}: R\&D on Lancium cloud integration/TCO
    \item [Cloud] From section \ref{Network}: R\&D in exploiting network interconnects and integrating cloud resources in LHCONE
    \item [HPC,Cloud] From section \ref{Accounting}: R\&D in accounting on opportunistic resources, particularly for workflows not run through a CE
    \item [Production systems] Expand production systems to support ML/AI training and optimization workflows
\end{itemize}

\appendix
\section{Charge} \label{charge}

\textbf{220301 - Charge for U.S. ATLAS / U.S. CMS commercial cloud/HPC Blueprint}

The Software \& Computing Operations Program managers of U.S. ATLAS and U.S. CMS would like to charge: \\

\noindent Lincoln Bryant (UC) \\
Fernando Barreiro (UTA) \\
Dirk Hufnagel (FNAL) \\
Kenyi Hurtado Anampa (Notre Dame) \\

to conduct a blueprint process about the usage of commercial cloud/HPC resources in the U.S. ATLAS and U.S. CMS collaborations. The blueprint is to conclude by 1. December 2022 with a document outlining findings and recommendations to the S\&C Operations Program managers of U.S. ATLAS and U.S. CMS. 

The report should cover:
\begin{itemize}
\item \textbf{[Workflows]}: What workflows (generation, simulation, reconstruction, etc.)  can be executed on what forms of commercial cloud/HPC resources? What metrics should be used to decide whether a workflow is executed efficiently, both in acquiring the resources and to operate the workflows on these resources?

\item \textbf{[Cost]}: What is the total cost of operating commercial cloud/HPC resources for collaboration workflows, both for computing resources as well as operational effort for the time ranges of LHC Run-3 and HL-LHC. Operational effort here is the work needed to run workflows on resources that had already been enabled.

\item \textbf{[Further R\&D]}: What further R\&D activities and/or development projects are needed to expand the range of  commercial cloud/HPC resources used in production and/or increase their cost effectiveness?

\item How is the usage of commercial cloud/HPC resources integrated into the experiment-wide infrastructure for workflow management and resource provisioning and resource utilization planning processes?

\item What can US ATLAS and US CMS recommend to the ATLAS and CMS collaborations to improve the utilization of commercial cloud/HPC resources, both in terms of enabling additional workflows as well as improving their cost effectiveness. 

\item What can US ATLAS and US CMS learn from related international efforts aimed at running collaboration workflows on commercial cloud/HPC resources? 

\item What new facility features or policies would help US ATLAS and US CMS adoption of commercial cloud/HPC resources? 

\item Additional topics/observations made during the blueprint process that should be mentioned in the final report?
\end{itemize}

The blueprint process should include a U.S. centric workshop as well as a later workshop with international participation. The final report should be the basis for publications and conference contributions. \\

Paolo Calafiura, Kaushik De, Tulika Bose, Oliver Gutsche

\section{Choice of Components in the Cloud} \label{cloud-componen-overview}
\subsection{Compute}

Major Cloud providers offer different pricing models with advantages and disadvantages that need to be understood.

%\textbf{Reserved instances} \\
\subsubsection{Reserved instances}
The customer commits to a certain amount of specific resources (e.g. a particular instance type) for a time frame usually between one and three years. List prices are further reduced for longer commitments, since it helps the cloud provider amortize the infrastructure investment. 

\begin{itemize}
    \item Advantages
    \begin{itemize}
        \item The reserved instances are guaranteed to be available for the customer.
        \item Prices are lower than on-demand instances, since the customer is paying the cost of the resources and the cloud provider runs no risk of purchasing unused resources.
    \end{itemize}
    \item Disadvantages  
        \begin{itemize}
        \item The resources are charged whether they are used or not. This limits the flexibility of cloud usage. 
        \item  The commitment is restricted to a certain Region and instance family (e.g. a particular CPU model).
    \end{itemize}
\end{itemize}

%\textbf{On demand instances} \\
\subsubsection{On demand instances}
The instances are requested on demand and billed by the minute or hour. There is absolutely no commitment from the customer. Once the customer has successfully claimed the instances, they are reserved for the customer until returned.
\begin{itemize}
    \item Advantages
    \begin{itemize}
        \item The customer does not make any commitment and uses the resources at his desire.
        \item The customer is not tied to any particular instance type or Region.
    \end{itemize}
    \item Disadvantages  
        \begin{itemize}
        \item These are the most expensive resources. 
        \item Cloud capacity is finite and there is no control over the activity of other customers (large corporations and organizations). There is therefore no guarantee that the desired amount of resources will be available. For instance, if ATLAS or CMS would want to run a large computation before an important physics conference, and the date collides with a particular online sales campaign, the amount of available resources might be limited. This is particularly true for less abundant chip families or architectures. The client should be flexible with Regions and instance types, and not corner themselves into a very popular product.
    \end{itemize}
\end{itemize}

\subsubsection{\textit{Spot} and \textit{preemptible} instances} 
These instances are similar to \textit{on-demand} instances, but can be claimed back by the cloud provider \textbf{with very short notice at any time}. \textit{Spot} and \textit{preemptible} instances come with an important price discount. These instances are not suitable to host critical applications or very long workloads, but can be very interesting for increasing the processing capacity and run inexpensive batch jobs. \textit{Preemptible} instances have a maximum lifetime (e.g. 24h in Google), while there is no predefined maximum lifetime on \textit{spot}.

\begin{itemize}
    \item Advantages
    \begin{itemize}
        \item The customer does not make any commitment and uses the resources at his pleasure.
        \item The customer is not tied to any particular instance type or Region.
        \item These are the least expensive instances.
    \end{itemize}
    \item Disadvantages  
        \begin{itemize}
        \item There is no guarantee from the Cloud provider that the desired amount of resources will be available. Due to their reduced price, these resources are claimed back whenever the cloud provider decides. During periods of high demand, the eviction rate of these instances increases. The customer needs to be careful relying exclusively on these instances for highly critical and urgent applications or workloads.
    \end{itemize}
\end{itemize}

\subsubsection{Disks} 
% Local SSD vs network mounted SSD VS HDD depending on workloads
% The disk price is always "on demand" and does not fall into Spot discounts
We will briefly discuss the disks mounted on the Virtual Machines. For batch jobs, this disk will store temporary files and is considered scratch disk, i.e. we do not need any longer term storage beyond the duration of the job. The two familiar options are persistent disks and local disks.

Persistent disks are durable \textbf{network} storage devices and are the default disks for any Virtual Machine. The data is distributed across several physical disks to ensure redundancy. Persistent disks are located independently from the Virtual Machine, so you can keep your data even after you delete an instances. Persistent disk throughput depends on the disk type (rotating disk or solid state disk) and scales with the size (you get higher throughput on a larger disk). Persistent disk throughput is throttled to the advertised value, in order to protect the installation and ensure all customers receive their contractual share. It is very important to validate the throughput characteristics of the disk matches the necessities of the application and select a more performant disk type or increase the size of the disk only to obtain a larger bandwidth. The throughput is generally low and the experiment's batch jobs do not require the advanced capabilities of persistent storage (redundancy or durability).

Local disks are \textbf{physically} attached to the server that hosts your VM instance. Local SSDs have much higher throughput than persistent disks. The data that you store on a local disk persists only until the instance is stopped or deleted. These disks are more suitable for experiment batch processing, however come at a higher cost, are not as widely available, i.e. you might not be able to scale out a large cluster with local disks, and the customer needs to maintain a certain virtual CPU to local disks proportion, e.g. 1 local SSD with 375 GB per each 8 vCPU.
The cost of the disk is added on top of the Virtual Machine CPU and memory and is always \textit{on-demand}. That means there are no discounts on disk for \textit{spot} or \textit{preemptible} instances.

\subsubsection{Takeaways}
Perhaps an important takeaway of this chapter is diversification, diversification and diversification:
\begin{itemize}
    \item Being aware of the capabilities of the different instance types and combining them according to the experiment's needs: e.g. having a baseline of reserved instances for critical applications/workloads and relying on \textit{on-demand} and \textit{spot} instances for flexible usage.
    \item Running on multiple Regions and Zones: running on different Regions and Zones increases the chances that the desired resources will be available.
    \item Using different instance families: being flexible on processor types increases the chance that one of them will be available. In most cases ATLAS and CMS don’t depend on exactly a particular CPU or GPU type and can fallback on others. 
    \item Using different instance sizes: while not ideal for large scale ups, smaller instance sizes are typically easier to allocate for the cloud provider. It is important to combine them for large clusters.
\end{itemize}

\subsection{Storage}
Cloud Storage is usually provided in different classes. The class names and exact characteristics depend on the provider. For example Google distinguishes Standard, Nearline, Coldline and Archive classes. Amazon offers General purpose, Unknown/Intelligent Tiering, Infrequent access and Archive tiers.

Excluding some of the archival classes, the access times and data availability are similar. The main difference is the pricing model. For \textbf{colder storage classes}, the main differences to watch are:
\begin{itemize}
    \item The minimum storage duration increases, i.e. you will be billed at least for the minimum duration (one or multiple months) even if you remove the data earlier. 
    \item You will be billed for the number of accesses.
\end{itemize}
The customer should be aware and have some statistics of his data access to choose the most adequate storage class.

 In practice we use the Standard or General purpose storage class for our active cloud sites, since data is often kept only for some days as an intermediate copy and will get accessed a few times during those days. 
 
 We did not evaluate long-term tape backup storage solutions as are present in the WLCG Tier1s. Cold or archival classes would be a better fit for these, but would require a deeper analysis to choose the most cost-efficient option. Archival storage in the cloud also raises the experiments' concerns over data custodiality and would require a more advanced risk management assessment.

It is also possible to automatically replicate the storage across multiple Regions in the same continent. Obviously the replication comes at a cost. Cloud-internal data replication would not fit transparently into the ATLAS computing model and therefore this feature has not been experimented with yet.

\subsection{Network}
It is important to have a high level overview of the network costs when using cloud resources, since it can add significantly to the bill. Transfers are billed by volume and the price depends on the distance between source and destination.
\begin{itemize}
    \item Ingress: data transfers \textbf{to} the cloud are free.
    \item Transfers in the cloud: data transfers in the cloud can incur a cost.
    \begin{itemize}
        \item Transfers between Availability Zones inside the same Region are usually free.
        \item Transfers between Regions: are usually at a cost. If the Regions are in different continents the cost will be higher, since it will need to go through expensive, intercontinental fibers.
    \end{itemize}    
    \item Egress: transferring data \textbf{out} of the cloud will be billed at a higher cost than the above cases. The egress cost can be significant depending on the activity and integration model. Egress is a common concern across experiments, mentioned at least by ATLAS, CMS and Vera Rubin. Cloud providers frequently offer 15\% of the compute bill in free egress, but this could not be enough to cover egress expenses. Experiments need to negotiate the terms for egress differently or modify their integration model to minimize the amount of data transferred out of the cloud.
\end{itemize}

% \bibliographystyle{}
% \bibliography{}

\end{document}